\documentclass[aip,graphicx,amsmath,amssymb,preprint]{revtex4-1}
\usepackage{CJKutf8}

\usepackage[capitalize]{cleveref}
\usepackage{graphics}
\usepackage{graphicx}
\usepackage{algorithm}
\usepackage{algorithmic}
\usepackage{xcolor}

\draft 

\begin{document}
\begin{CJK*}{UTF8}{gbsn}

\title{Synthetic turbulence generator for lattice Boltzmann method at the interface between RANS and LES} 



\author{Xiao Xue(薛骁)}
 \email{xiaox@chalmers.se}
\author{Hua-Dong Yao(姚华栋)}
%
\author{Lars Davidson}%
\affiliation{ 
Division of Fluid Dynamics, Department of Mechanics and Maritime Sciences, Chalmers University of Technology, 41296, Gothenburg, Sweden
}%


\date{\today}

\begin{abstract}
The paper presents a synthetic turbulence generator (STG) for the lattice Boltzmann method (LBM) at the interface of the Reynolds averaged Naiver-Stokes (RANS) equations and the LBM Large Eddy Simulation (LES). We first obtain the RANS velocity field from a finite volume solver at the interface. Then, we apply a numerical interpolation from the RANS velocity field to the LBM velocity field due to the different grid types of RANS and LBM. The STG method generates the velocity fluctuations, and the regularized LBM reconstructs the particle distribution functions at the interface. We perform a turbulent channel flow simulation at $\mathrm{Re}_{\tau} = 180$ with the STG at the inlet and the pressure-free boundary condition at the outlet. The velocity field is quantitatively compared with the periodic lattice Boltzmann based LES (LES-LBM) channel flow and the direct numerical simulation (DNS) channel flow. Both adaptation length and time for the STG method are evaluated. Also, we compare the STG-LBM channel flow results with the existing LBM synthetic eddies method (SEM-LBM) results. Our numerical investigations show good agreement with the DNS and periodic LES-LBM channel flow within a short adaptation length. The adaptation time for the turbulent channel flow is quantitatively analyzed and matches the DNS around 1.5 to 3 domain flow-through time. Finally, we check the auto-correlation for the velocity components at different cross-sections of the streamwise direction. The proposed STG-LBM is observed to be both fast and robust. The findings show good potential for the hybrid RANS/LES-LBM based solver on the aerodynamics simulations and a broad spectrum of engineering applications. (This paper has been accepted as a journal paper in Physics of Fluids)
\end{abstract}

\maketitle 

\section{INTRODUCTION}
In computational fluid dynamics (CFD), there are rising needs to understand the detailed characteristics of turbulence ranging from fundamental to real-world applications. To tackle these problems, direct numerical simulation (DNS) and large eddy simulation (LES)~\cite{smagorinsky1963general, spalart1992one} are frequently used to resolve flow structures. However, they are suffering from high computational cost. Reynolds averaged Navier-Stokes (RANS) simulation gives hope of low computational cost and reliable physics~\cite{menter1993zonal}. When detailed flow information is needed to resolve, LES is still necessary. Hybrid RANS/LES approach offers an excellent opportunity to save considerable computational cost, whereas it preserves detailed flow information. This requires correct modeling of the RANS/LES interface by capturing velocity fluctuations with little spurious noise~\cite{wu2017inflow}. One approach is to use precursor DNS or LES data. However, this approach is limited to fundamental flows studies with simple configurations and low Reynolds numbers~\cite{schluter2004large}.
Another approach is "recycling" the velocity field back to the inlet of the simulation~\cite{lund1998generation, spalart2006direct, shur2011rapid}. This method is still constrained by the flow complexity when a strong pressure gradient occurs. The synthetic eddy method (SEM) can handle more general hybrid RANS/LES cases but suffers from a relatively long adaptation length for the fully developed turbulence~\cite{mathey2006assessment, jarrin2009reconstruction, skillen2016accuracy}. By introducing dynamic control forcing techniques (DCFT) to the SEM framework, Roidl et al.~\cite{roidl2011zonal, roidl2012zonal} showed good performance of the case for a transonic airfoil with an adaptation region of merely 2-3 boundary-layer widths. However, the DCFT is computationally expensive, which limits its application bandwidth. Synthetic turbulence generators give opportunities to balance between accuracy, ease of implementation and computationally relatively cheap~\cite{klein2003digital, di2006synthetic, davidson2007using, xie2008efficient, huang2010general, shur2014synthetic}. Shur et al.~\cite{shur2014synthetic} highlighted the STG method in aerodynamics and acoustics applications with a considerably short adaptation length. 

Knowing that the LES is computationally expensive using conventional CFD due to the data dependencies, the lattice Boltzmann method (LBM) can dramatically improve the computational efficiency since the information on each lattice cell is stored and updated locally. Unlike conventional CFD methods, each cell in LBM does not need to wait for its neighbors to update during the algorithm's evolution~\cite{succi2001lattice, kruger2017lattice, marie2009comparison}. It is worth mentioning that LBM is not solving the macroscopic scale quantities directly like Navier-Stokes equations. Governed by the Boltzmann equation, LBM computes the particle's distribution function in mesoscopic scales. Through the Chapman-Enskog procedure, LBM can recover the Navier-Stokes equations. In the past decades, LBM has become widespread on account of its broad applicability across a range of  complex fluid dynamics problems ranging from micro-nano scales\cite{Belardinelli15, xue2020brownian, xue2018effects} to macroscopic scales\cite{hou1994lattice, toschi2009lagrangian, karlin1999perfect, lallemand2000theory} at low Mach numbers. They have successfully studied complex fluid phenomena including turbulence\cite{teixeira1998incorporating,chen1998lattice,karlin1999perfect} and non-ideal fluids with phase transition and/or segregation\cite{he1999lattice, liu2012three, reis2007lattice, chiappini2019, chiappini2018ligament,xue2018effects}. Hou et al.~\cite{hou1994lattice} brought the Smagorinsky LES into the lattice Boltzmann models by introducing the effective viscosity. This contribution enables LBM to improve the computational efficiency for LES computation. Hybrid RANS and LES-LBM method can further reduce the computational cost of complex engineering applications. Thus, it is crucial to impose turbulence in the LES-LBM framework accurately and efficiently. Although it is an active area for the traditional CFD community, it is still an early-stage development for the lattice Boltzmann method. Previous studies have used passive method by putting obstacles in the flow field and removing obstacles after the turbulence is triggered~\cite{koda2015lattice}. However, this method has a relatively long adaptation time.
Fan et al.~\cite{fan2021source} have extended the SEM method in the lattice Boltzmann method by reconstructing the force. However, it takes 15 boundary layer lengths to match DNS results fully. Buffa et al.~\cite{buffa2021lattice} have successfully adapted the SEM method into the LBM framework by reconstructing the velocity field into particle distribution functions. The above methods only consider the LES implementation without considering the hybrid RANS input. In the present work, we integrate the synthetic turbulent generator~\cite{shur2014synthetic} into the LBM framework(STG-LBM) so that it can be used at a hybrid RANS/LES-LBM interface. The STG-LBM shows relatively short adaptation length with only 2-4 boundary-layer thicknesses and fast converging speed with around 1.5-3 domain flow-through time.\\ This paper is organized as follows:~\cref{sec:methodology} introduces the lattice Boltzmann method with the Bhatnagar–Gross–Kroog (BGK) collision operator and its Smagorinsky subgrid-scale (SGS) LES modeling~\cite{hou1994lattice, smagorinsky1963general}. In~\cref{sec:num-stg}, we review the STG method. Then, we describe how to integrate the synthetic turbulent generator to the LBM framework at the RANS/LES-LBM interface.~\cref{sec:set-up} shows the turbulent channel flow set up with the hybrid RANS/LES method. In~\cref{sec:results}, we simulate the turbulent channel flow at $\mathrm{Re}_{\tau} = 180$. The results compare with the DNS, LES-LBM periodic channel flow, and the SEM-LBM channel flow to show the convergence speed of the present method in both space and time. Also, we examine the auto-correlations at the different cross-sections of the channel flow. The conclusions are summarized in~\cref{sec:conclusions}.

\section{METHODOLOGY}\label{sec:methodology}
In this section, we first present the D3Q19 lattice Boltzmann method with the single relaxation time scheme. Then, the Smagorinsky subgrid-scale modeling for LES in the lattice Boltzmann framework is presented. 

\subsection{The lattice Boltzmann method}\label{sec:method-lbm}
The lattice Boltzmann method is a dimensionless model with all quantities formed in the dimensionless lattice Boltzmann units(LBU). We apply a three dimensional(3D) lattice model called the D3Q19 model which has 19 discretized velocity directions $\mathbf{c}_i$ ($i=0...Q-1$). The particle's probability distribution function, $f_{i}(\mathbf{x},t)$, denotes the $i$-th direction of a lattice cell. The macro-scale quantities for the density, $\rho(\mathbf{x}, t)$, momentum, $\rho(\mathbf{x}, t)\mathbf{u}(\mathbf{x}, t)$, and momentum flux tensors, $\mathbf{\Pi}(\mathbf{x}, t)$, can be calculated from the distribution function, the discrete velocities, and the volume force:
\begin{equation}\label{eq:density}
\rho(\mathbf{x}, t) = \sum_{i=0}^{Q-1} f_i(\mathbf{x}, t), \\
\end{equation} 
\begin{equation}
\label{eq:momentum}
\rho(\mathbf{x}, t)\mathbf{u}(\mathbf{x}, t) = \sum_{i=0}^{Q-1} f_i(\mathbf{x}, t)\mathbf{c}_{i} + \frac{1}{2}\mathbf{F}\Delta t,
\end{equation}
\begin{equation}
\label{eq:tensor}
\mathbf{\Pi}(\mathbf{x}, t) = \sum_{i=0}^{Q-1} f_i(\mathbf{x}, t)\mathbf{c}_{i}\mathbf{c}_{i},
\end{equation}
where the momentum flux, $\mathbf{\Pi}$, can be presented by the sum of the equilibrium and non-equilibrium parts, $\mathbf{\Pi}(\mathbf{x}, t)  = \mathbf{\Pi}_{\text{eq}} (\mathbf{x}, t)  + \mathbf{\Pi}_{\text{neq}}(\mathbf{x}, t)$. $\Delta t$ is the marching time step, which is set to unity in the LBM algorithm.
The lattice cell is located at position $\mathbf{x}$ at time $t$. The evolution equation for the distribution functions, considering collision and forcing, can be written as:
\begin{equation}
\label{eq:lbe}
f_i(\mathbf{x}+\mathbf{c}_{i}\Delta t,t+\Delta t) =f_i(\mathbf{x}, t) + \Omega\left[f_i(\mathbf{x},t )-f_i^{\mbox{ eq}}(\mathbf{x},t )\right] + \Delta t F_i(\mathbf{x}, t),\\
\end{equation}
where $\Omega$ is a collision kernel, we use the Bhatnagar–Gross–Kroog (BGK) collision kernel which has been widely adopted to various applications~\cite{succi2001lattice}, with $\Omega = \frac{\Delta t}{\tau}$. The BGK collision operator fixes the single relaxation time $\tau$ for the colliding process. The collision kernel relaxes the distribution function towards the local Maxwellian distribution function $f_i^{\mbox{ eq}}$~\cite{succi2001lattice, kruger2017lattice}:
\begin{equation}
\label{eq:local_eq}
f_i^{\mbox{ eq}}\left(\textbf{x},t\right) =  \omega_{i}\rho\left(\textbf{x},t\right) \bigg[1+\frac{\textbf{c}_i\cdot\textbf{u}\left(\textbf{x},t\right)}{c_s^2}+\frac{\left[\textbf{c}_i\cdot\textbf{u}\left(\textbf{x},t\right)\right]^2}{2c_s^4}  - \frac{\left[\textbf{u}\left(\textbf{x},t\right)\cdot\textbf{u}\left(\textbf{x},t\right)\right]}{2c_s^2}\bigg] \; ,
\end{equation}
where $c_s$ is the speed of the sound which is equal to $c_s = 1/\sqrt{3}$ LBU. The kinematic viscosity can be calculated by:
\begin{equation}
\label{eq:lbe-nu}
\nu = c_s^2(\tau-\frac{1}{2})\Delta t.
\end{equation}
$F_i$ in~\cref{eq:lbe} is the volume force acting on the fluid following the approach from Guo~\cite{guo2002discrete} which can be obtained by:
\begin{equation}
\label{eq: bodyforce}
F_i(\mathbf{x}, t) = (1 - \frac{1}{2\tau}) \omega_i \left [ \frac{\textbf{c}_i - \textbf{u}\left(\textbf{x},t\right)}{c_s^2} + \frac{\textbf{c}_i \cdot \textbf{u}\left(\textbf{x},t\right)}{c_s^4}\textbf{c}_i \right] \mathbf{F},
\end{equation}
where $\mathbf{F}$ is the volume acceleration.

\subsection{Smagorinsky subgrid-scale modeling}\label{sec:method-sgs}
Below, we recall the basic formulation of the Smagrinsky SGS large-eddy simulations techniques using the lattice Boltzmann method. Interested readers can refer to the literature~\cite{hou1994lattice, smagorinsky1963general, koda2015lattice}. The key of the Smagorinsky SGS modeling is to model the effective viscosity $\nu _{\text{eff}}$, which can be seen as the sum of the molecular viscosity $\nu_0$ and the turbulent viscosity $\nu_t$:
\begin{equation}
\nu _{\text{eff}}=\nu _0+\nu_t, \hspace{.6in} \nu_t = C\delta ^2\left |\bar{\mathbf{S}}\right |,
\label{smogrinsky_model}
\end{equation}
where $C$ is the Smagorinsky constant, $\delta$ represents the filter size, and $\left |\mathbf{\bar{S}} \right |$ is the filtered strain rate tensor:
\begin{equation}
\label{eq: strain}
\left |\bar{\mathbf{S}}\right | = \frac{-\tau_0 \rho c \Delta x + \sqrt{(\tau_0 \rho c \Delta x)^2 + 18\sqrt{2}\rho C \delta^2 Q^{1/2}}}{6 \rho C \delta^2},
\end{equation}
where $\tau _0$ is the original relaxation time from the BGK collision operator, $c = \Delta x / \Delta t$, and $Q^{1/2}$ can be written as:
\begin{equation}
Q^{1/2} = \sqrt{\mathbf{\Pi}_{\text{neq}}:\mathbf{\Pi}_{\text{neq}}},
\end{equation}
where $\mathbf{\Pi}_{\text{neq}}$ is the non-equilibrium part of the momentum flux tensor $\mathbf{\Pi}$ shown in~\cref{eq:tensor}. With help of~\cref{eq:lbe-nu}, we can transfer the effective viscosity into the effective relaxation time. Following \cite{koda2015lattice}, we can obtain the total relaxation time $\tau _{\text{eff}}$, which is written as:
\begin{equation}
\label{eq:tau_eff}
    \tau _{\text{eff}}=\frac{\tau _0}{2}+\frac{\sqrt{(\tau_0\rho c)^2 + 18\sqrt{2}C Q^{1/2}}}{2\rho c}.
\end{equation}
Finally, we replace $\tau$ in the BGK collision operator with $\tau _{\text{eff}}$ to enclosure the lattice Boltzmann based LES system.

\section{Synthetic turbulence generator in the Lattice Boltzmann framework}\label{sec:num-stg}
The synthetic turbulence generator (STG) has been used at the interface of RANS/LES or RANS/IDDES~\cite{shur2014synthetic} by using the FVM. In the LBM community, previous studies have only integrated the Synthetic Eddy Method (SEM)~\cite{buffa2021lattice, fan2021source} into the lattice-Boltzmann-based framework. Notice that STG is different from the SEM method. In the SEM, coherent structures are simulated by superimposing artificial eddies at the inlet plane. Each eddy is given vorticity with a three-dimensional structure represented by a shape function describing the spatial and temporal characteristics of the structure. As for the STG method, the synthetic fluctuations are created using a Fourier series where the Fourier coefficient is given by the $-5/3$ energy spectrum. In the present work, we mainly focus on integrating the STG into the hybrid RANS/LES-LBM framework. In the following, we will briefly remind users about the STG method and show how we couple RANS with the LES-LBM using the STG method. Finally, the particle distribution functions at the interface is computed from the velocity field via the regularized lattice Boltzmann method~\cite{Latt2007}.
\subsection{Synthetic turbulence generator formulation}\label{sec:method-stg}
The synthetic turbulence generator at the interface between the RANS simulation and the LES simulation reconstructs the velocity  $\mathbf{u}(\mathbf{x}, t)$ at the cell $\mathbf{x}$ at time $t$:
\begin{equation}
\label{eq:u_couple}
\mathbf{u}(\mathbf{x}, t) = \mathbf{u}(\mathbf{x})_{\text{RANS}} + \mathbf{u}'(\mathbf{x}, t),
\end{equation}
where $\mathbf{u}(\mathbf{x})_{\text{RANS}}$ is the velocity vector obtained from the RANS results, and $\mathbf{u}'(\mathbf{x}, t)$ is the vector of the velocity fluctuations. 
The time-averaged fluctuation is zero, i.e. $\langle \mathbf{u}'(\mathbf{x}, t)\rangle = 0$. The velocity component from $\mathbf{u}'(\mathbf{x}, t)$ can be expressed  as $u_{\alpha}(\mathbf{x}, t)$:
\begin{equation}
\label{eq:v_fluc}
\mathbf{u}'(\mathbf{x}, t)=a_{\alpha\beta}\mathbf{v}'(\mathbf{x}, t),
\end{equation}
where $a_{\alpha\beta}$ is the Cholesky decomposition of the Reynolds stress tensor:
\begin{equation}
\label{eq:cholesky}
\left \{  a_{\alpha\beta}\right \}  = \begin{pmatrix}
\sqrt{R_{11}} & 0 & 0\\ 
R_{21}/a_{11} & \sqrt{R_{22}-a_{21}^2} & 0\\ 
R_{31}/a_{11} & (R_{32}-a_{21}a_{31})/a_{22} & \sqrt{R_{33}-a_{31}^2-a_{32}^2}
\end{pmatrix},
\end{equation}
where $R_{\alpha\beta} = \left\langle u'_\alpha u'_\beta \right\rangle$ is taken from Reynolds stresse tensor of the RANS simulation.
The auxiliary vector of the velocity fluctuations, $\mathbf{v}'(\mathbf{x}, t)$, are constructed by superposition of N weighted Fourier modes
\begin{equation}
\label{eq:aux_vec} 
\mathbf{v}'(\mathbf{x}, t)  = \sqrt{6} \sum_{n=1}^{N}\sqrt{q^n}\left [ \mathbf{\sigma}^n \mathrm{cos}\left ( k^n\mathbf{d}^n\cdot\mathbf{x}' + \phi^n  \right ) \right ],
\end{equation}
where $n$ denotes the mode number, $\mathbf{d}^n$ is the random unit vector that is uniformly located over a sphere which is
\begin{equation}
    \label{eq:dn}
    \mathbf{d}^n=\begin{pmatrix}
\mathrm{sin}(\Theta)\mathrm{cos}(\Phi)\\ 
\mathrm{sin}(\Theta)\mathrm{sin}(\Phi)\\ 
\mathrm{cos}(\Theta)
\end{pmatrix},
\end{equation}
where $\Phi$ is uniformly distributed in the interval of $[0, 2\pi)$ and $\Theta = \mathrm{arccos}(1-\gamma/0.5)$ where $\gamma$ is uniformly distributed in the interval of $[0, 2\pi)$. $k^n$ is the amplitude of the vector $\mathbf{d}^n$, $\mathbf{\sigma}^n$ represents the unit vector normal to the vector $\mathbf{d}^n$($\mathbf{\sigma}^n \cdot \mathbf{d}^n = 0$)~\cite{davidson2018fluid, kraichnan1970diffusion} which can be written as 
\begin{equation}
    \label{eq:dn}
    \mathbf{\sigma}^n=\begin{pmatrix}
        \mathrm{cos}(\Phi)\mathrm{cos}(\Theta)\mathrm{cos}(\eta)-\mathrm{sin}(\Phi)\mathrm{sin}(\eta)\\ 
        \mathrm{sin}(\Phi)\mathrm{cos}(\Theta)\mathrm{cos}(\eta)+\mathrm{cos}(\Phi)\mathrm{sin}(\eta)\\ 
        -\mathrm{sin}(\Theta)\mathrm{cos}(\eta)
    \end{pmatrix},
\end{equation}
where $\eta$ is random number that uniformly distributed in the interval of $[0, 2\pi)$. $\phi^n$ is random number that uniformly distributed in the interval of $[0, 2\pi)$. $q^n$ represents the normalized amplitude of a modified von Karman spectrum~\cite{shur2014synthetic}:

\begin{equation}
\label{eq:qn}
q^n = \frac{E(k^{n})\Delta k^n}{\sum^{N}_{n=1}E(k^{n})\Delta k^n}, \hspace{.6in} \sum^{N}_{n=1} q^n = 1,
\end{equation}
where $E(k^{n})$ is the modified von Karman spectrum. $\mathbf{x}'$ in~\cref{eq:aux_vec} is the pseudo-position vector which can be defined as

\begin{equation}
\label{eq:psu}
\mathbf{x}'=\begin{pmatrix}
\frac{2\pi }{k^n \mathrm{max}\left\{l_e(\mathbf{x})\right\}}(x - U_{b}t)\\ 
y\\ 
z
\end{pmatrix}.
\end{equation}
In~\cref{eq:psu}, $U_{b}$ is the bulk velocity at the RANS/LES interface, and $l_e(\mathbf{x})$ is the local scale of the most energy-containing eddies. $l_e(\mathbf{x})$ is given by $l_e(\mathbf{x})=\mathrm{min}(2d_w(\mathbf{x}), C_ll_t)$, where $d_w(\mathbf{x})$ is the distance of the cell to the wall. $C_l$ is the empirical constant $C_l = 3.0$, $l_t$ is the length-scale of the RANS model($l_t=\sqrt{k_t}/0.09\omega_t$), where $k_t$ and $\omega_t$ are given by the $k-\omega$ RANS model~\cite{wilcox:88}. In~\cref{eq:aux_vec}, $N$ mode is obtained by 
\begin{equation}
\label{eq:Nmode}
N=\mathrm{ceil}\left \{\mathrm{ln}(k_{\text{max}} / k_{\text{min}}) / \mathrm{ln}(1 + \alpha) + 1)\right \},    
\end{equation}
where $\alpha = 0.01$, $k_{\text{min}}=\pi/\mathrm{max}\left \{l_e(\mathbf{x})\right \}$, and $k_{\text{max}}=1.5\mathrm{max}\left \{2\pi/l_{\text{cut}}\right \}$. $l_{\text{cut}}$ is given by
\begin{equation}
\label{eq:lcut}
l_{\text{cut}}=2\mathrm{min}\left \{ \mathrm{max}\left\{d_x, d_y, 0.3d_{\text{max}}\right \}+0.1d_w, d_{\text{max}}\right \},
\end{equation} 
where $d_x, d_y, d_z$ are the local grid size of the RANS simulation at the RANS/LES-LBM interface, $d_{\text{max}} = \mathrm{max}\left\{d_x, d_y, d_z\right \}$. $\mathbf{v}'$ is satisfying $\left\langle v'_{\alpha}v'_{\beta}\right\rangle = 0$, $\left\langle v'_{\alpha\beta} \right\rangle = \delta_{\alpha\beta}$ ($v'_{\alpha}$ is the component of $\mathbf{v}'$). We have now reviewed the STG approach in the macroscopic CFD solver for the RANS/LES interface. However, LBM is a dimensionless mesoscopic approach. Direct coupling from RANS to LES-LBM is non-straightforward. Below, we demonstrate how to impose the turbulent from the RANS/LES-LBM interface.

\subsection{Boundary reconstruction with the regularized lattice Boltzmann method}\label{sec:lbm-regularized}
\begin{figure}[t]
\centering
\includegraphics[width=0.35\textwidth]{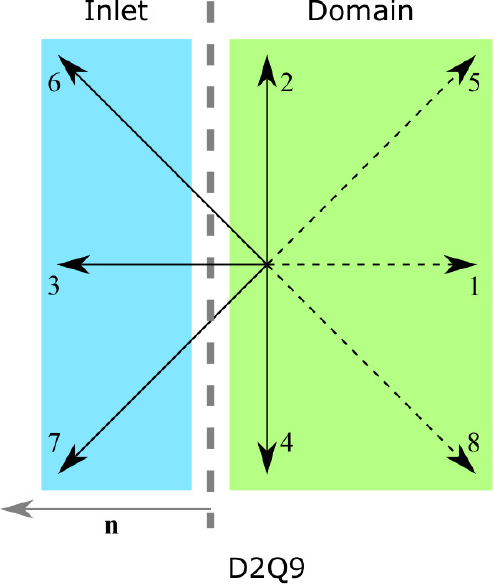}
\caption{The sketch of the D2Q9 lattice model at the inlet boundary. The dotted lines are unknowns. The blue part is the inlet boundary, whereas the green part is the computational domain. $\mathbf{n}$ is the unit normal vector that is perpendicular the to boundary pointing towards the inlet.} 
\label{fig:sketch_D2Q9_inlet}
\end{figure}

The particle's probability distribution function at the interface, $f_i^{\text{bc}}$, can be defined as
\begin{equation}
\label{eq:f_i}
f^{\text{bc}}_i(\mathbf{x}, t) = f^{\text{bc(eq)}}_i(\mathbf{x}, t) + f^{\text{bc(neq)}}_i(\mathbf{x}, t),
\end{equation}
where $f^{\text{bc(neq)}}_i(\mathbf{x}, t)$ denotes the non-equilibrium part of the distribution function. The superscript "bc" denotes the boundary condition at the interface. The present work applies the regularized scheme proposed by Latt, et al.~\cite{latt2008straight} to reconstruct the non-equlibrium part of population which can be calculated via
\begin{equation}
\label{eq:f_i_neq}
f^{\text{bc(neq)}}_i(\mathbf{x}, t)\approx \frac{\omega_i}{c_s^4} \mathbf{Q}_i:\mathbf{\Pi}^{\text{bc}}_{\text{neq}},
\end{equation}
where $\mathbf{Q}_i = \mathbf{c}_i\mathbf{c}_i - c_s^2\mathbf{I}$ with $\mathbf{I}$ being the identity matrix. $\mathbf{\Pi}^{\text{bc}}_{\text{neq}}$ is the non-equilibrium part of the moment flux tensor which is defined as
\begin{equation}
\label{eq:pi_neq}
\mathbf{\Pi}_{\text{neq}} = \sum^{Q-1}_{i=0}\mathbf{Q}_if^{\text{bc(neq)}}_i(\mathbf{x}, t).
\end{equation}
Note that, at the boundary of the domain there are unknown variables in the $i$ the directions,  which can be calculated via the known direction following:
\begin{equation}
\label{eq:unknown_neq}
\mathbf{Q}_i = \mathbf{\bar{Q}}_{\text{inv}(i)},\hspace{.6in} f^{\text{bc(neq)}}_i(\mathbf{x}, t) = \bar{f}^{\text{bc(neq)}}_{\text{inv}(i)}(\mathbf{x}, t),
\end{equation}
where $\mathbf{\bar{Q}}_{\text{inv}(i)}$ and $\bar{f}^{\text{bc(neq)}}_{\text{inv}(i)}(\mathbf{x}, t)$ are the opposite direction of the unknown direction.
For example, ~\cref{fig:sketch_D2Q9_inlet} shows the case of the D2Q9 LBM model, where the unknown directions(1, 5 and 8) can be calculated by the opposite direction (3, 7 and 6) respectively; the same logic applies to the D3Q19 lattice model. We would like to comment that this method could also adapt to other 3D lattice models, for instance, D3Q15 and D3Q27. For the density of the inlet boundary, we follow the idea from Zou and He~\cite{zou1997pressure}.
\begin{equation}
\label{eq:density_bc}
\rho_{\text{bc}}(\textbf{x},t) = \frac{1}{1+\hat{u}_{\text{bc}}(\mathbf{x}, t)}(2\rho_{\bot}(\textbf{x},t) + \rho_{\parallel}(\textbf{x},t)),
\end{equation}
where $\hat{u}_{\text{bc}}$ is the cross product with the normal unit vector $\mathrm{n}$ at the boundary $\hat{u}_{\text{bc}}=\mathbf{u}^{\text{bc}}_{\text{LB}}\cdot \mathbf{n}(\left | \hat{u}_{\text{bc}} \right |<0.3c_s)$ and $\mathbf{u}^{\text{bc}}_{\text{LB}}$ is the velocity of the lattice Boltzmann domain at the interface. $\rho_{\bot}$ and $\rho_{\parallel}$ are the density calculated by
\begin{equation}
\label{eq:density_bc_sub}
\rho_{\bot}(\textbf{x},t)=\sum_{i\in\left\{ i| \mathbf{c}_i \cdot \mathbf{n} = 0\right\}}f^{\bot}_i(\mathbf{x}, t),\hspace{.6in}\rho_{\parallel}(\textbf{x},t) = \sum_{ i\in\left\{ i| \mathbf{c}_i \cdot \mathbf{n} < 0\right\}}f^{\parallel}_i(\mathbf{x}, t),
\end{equation}
where $\mathbf{n}$ is the normal vector pointing towards the boundary, $f^{\bot}_i$ and $f^{\parallel}_i$ are the probability density functions that point towards the boundary and are parallel to the boundary. In~\cref{fig:sketch_D2Q9_inlet}, $f^{\bot}_i$ is represented by direction 3, 6, 7, and $f^{\parallel}_i$ is represented by direction 2 and 4. Take~\cref{eq:density_bc} into~\cref{eq:local_eq}, the equilibrium distribution function at the boundary can be defined as
\begin{equation}
\label{eq:bc_eq}
f^{\text{bc(eq)}}_i\left(\textbf{x},t\right) =  \omega_{i}\rho_{\text{bc}}(\textbf{x},t) \bigg[1+\frac{\textbf{c}_i\cdot\mathbf{u}^{\text{bc}}_{\text{LB}}\left(\textbf{x},t\right)}{c_s^2}+\frac{\left[\textbf{c}_i\cdot\mathbf{u}^{\text{bc}}_{\text{LB}}\left(\textbf{x},t\right)\right]^2}{2c_s^4}  - \frac{\left[\mathbf{u}^{\text{bc}}_{\text{LB}}\left(\textbf{x},t\right)\cdot\mathbf{u}^{\text{bc}}_{\text{LB}}\left(\textbf{x},t\right)\right]}{2c_s^2}\bigg] \;,
\end{equation}
In~\cref{sec:method-rans}, we describe how to calculate the LBM velocity field, $\mathbf{u}^{\text{bc}}_{\text{LB}}$, from the RANS velocity field, $\mathbf{u}^{\text{bc}}_{\text{RANS}}$.
\begin{figure}[h]
\centering
\includegraphics[width=1\textwidth]{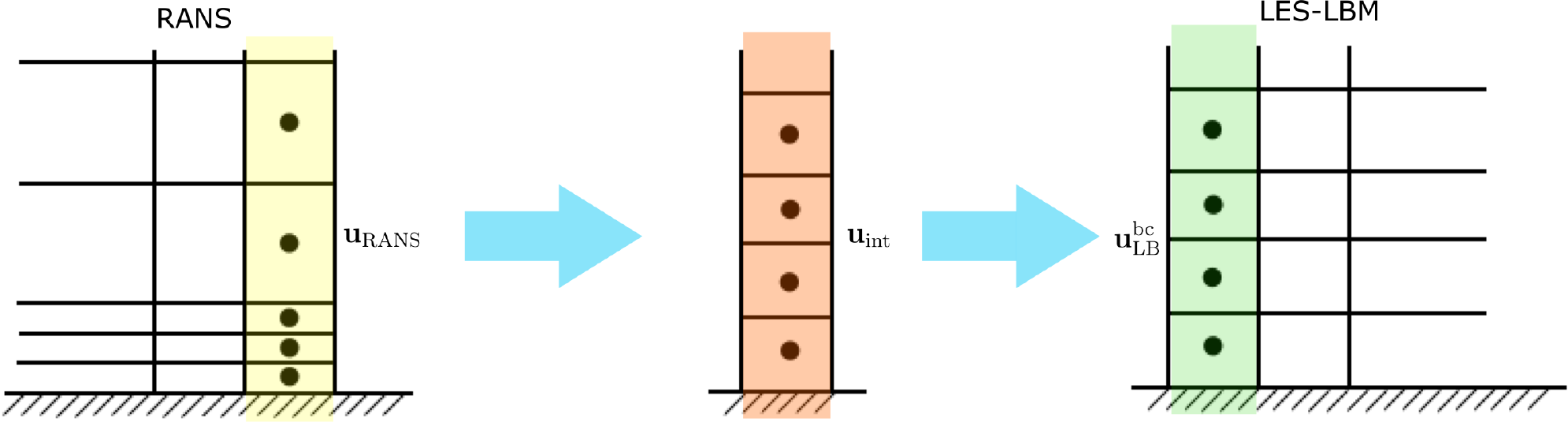}
\caption{Overview of how to couple the RANS with the LES-LBM framework. The left panel is the RANS computational domain. The colored layer is the velocity field at the RANS/LES-LBM interface. The middle part is the interpolated velocity field obtained from the RANS data. The right panel shows the sketch of the LES-LBM computational domain. The inlet velocity field (colored light green) is reconstructed from the interpolated RANS mean profile with superimposed synthetic fluctuations.} 
\label{fig:sketch_RANS/LBM}
\end{figure}
\subsection{Integration of synthetic turbulence generator at RANS/LES-LBM interface}\label{sec:method-rans}
At the hybrid RANS/LES-LBM interface,  we first compute the RANS flow field in a channel with the $k-\omega$ model~\cite{wilcox:88} using a finite-volume method (FVM). Thus, we can obtain the physical macroscale flow quantities like
the velocity, $k$ and $\omega$. The lattice Boltzmann model is based on a uniform Cartesian grid and the RANS FVM calculation is based on a non-uniform Cartesian grid. Therefore, it is necessary to interpolate the RANS velocity, $\mathbf{u}_{\text{RANS}}$, to the uniform grid velocity, $\mathbf{u}_{\text{int}}$, at the interface which is the $\mathbf{u}(\mathbf{x})_{\text{RANS}}$ in~\cref{eq:u_couple}. Second, we calculate the interpolated velocity fluctuations imposed at the interface following
\begin{equation}
\label{eq:u_phys}
\mathbf{u}^{\text{bc}}_{\text{phys}}(\mathbf{x}, t) = \mathbf{u}_{\text{int}}(\mathbf{x}) + \mathbf{u}'(\mathbf{x}, t),
\end{equation}
where $\mathbf{u}{'(\mathbf{x}, t)}$ is generated by the STG appoarch and can be obtained by~\cref{eq:v_fluc}-\cref{eq:psu}.\\
Now, we can calculate the dimensionless velocity field, $\mathbf{u}^{\text{bc}}_{\text{LB}}(\mathbf{x}, t)$, at the LES-LBM inlet by
\begin{equation}
\label{eq:u_transfer}
\mathbf{u}^{\text{bc}}_{\text{LB}}(\mathbf{x}, t) = \mathbf{u}^{\text{bc}}_{\text{phys}}(\mathbf{x}, t) \frac{c_t}{c_x},
\end{equation}
where $c_x$ and  $c_t$ are the conversion factor from the lattice Boltzmann simulation to physical system which is defined as
\begin{equation}
\label{eq:c_factor}
c_x = \frac{L_{\text{phys}}}{L_{\text{LB}}},\hspace{.6in} c_t = \frac{t_{\text{phys}}}{t_{\text{LB}}}, 
\end{equation}
where $L_{\text{phys}}$ and $t_{\text{phys}}$ represent the space and time unit from the physical system, $L_{\text{LB}}$ and $t_{\text{LB}}$ are the space and time lattice Boltzmann unit from the lattice Boltzmann simulation.
Following~\cref{eq:f_i}-~\cref{eq:density_bc_sub}, we use the regularized lattice Boltzmann method to reconstruct the particle distribution function $f_i(\mathbf{x}, t)$ at the inlet boundary with correct velocity and density quantities. Finally, we apply the streaming and collision step to updating the fluid field. ~\cref{fig:sketch_RANS/LBM} shows the general process of integrating the STG method at the RANS/LES-LBM interface.
\subsection{Implementation summary}\label{sec:method-sgs}
Now, we have formulated all steps. The short summary of the implementation and the formulas of the synthetic turbulence generator at the RANS/LES-LBM interface is presented in the following pseudo-code.
\begin{algorithm}[H]
   \caption{Implementation of synthetic turbulence generator for the RANS/LES-LBM interface (STG-LBM)}
   \label{alg:stg-lbm}
   \begin{algorithmic}
    \STATE \textbf{RANS section}
    \STATE 1. Perform a RANS simulation with a finite volume method using the $k-\omega$ model.
    \STATE 2. Save the Reynolds stress tensor $R_{\alpha\beta}$, the velocity field $\mathbf{u}_{\text{RANS}}$, the bulk velocity  $U_{b}$, $k$ and $\omega$ and other data needed for computing the velocity fluctuations $\mathbf{u}'(\mathbf{x}, t)$.
    \STATE \textbf{LBM section}
    \STATE 1. Obtain the RANS velocity at the interface $\mathbf{u}(\mathbf{x})_{\text{RANS}}$ in~\cref{eq:u_couple}.
    \STATE 2. Read the other saved value of $R_{\alpha\beta}$, the bulk velocity $U_{b}$, $k$ field, $\omega$ field, and other data from the RANS simulation.
    \STATE 3. Calculate $a_{\alpha\beta}$ using the Cholesky decomposition of the Reynolds stress tensor~\cref{eq:cholesky}.
    \STATE 4. Calculate the normalized amplitude $q^n$ according to~\cref{eq:qn}.\\
    \textbf{for all} $t=0$ till $t=t_{end}$ \textbf{do}\\
        \hskip1.0em \textbf{for all} cells \textbf{do}\\
            \hskip2.0em \textbf{if} cell $\mathbf{x}$ is at the RANS/LBM interface \textbf{then}\\
                \hskip3.0em 5. Update the pseudo-position $\mathbf{x}'$ according to~\cref{eq:psu}.\\
                \STATE \hskip3.0em  6. Calculate $\mathbf{v}'(\mathbf{x}, t)$ thanks to~\cref{eq:aux_vec},~\cref{eq:Nmode} and~\cref{eq:lcut}.\\
                \STATE \hskip3.0em  7. Compute the fluctuating velocity vector $\mathbf{u}'(\mathbf{x}, t)$ and $\mathbf{u}^{\text{bc}}_{\text{phys}}$ following~\cref{eq:v_fluc} and~\cref{eq:u_phys} respectively.\\
                \STATE \hskip3.0em  8. Compute $\mathbf{u}^{\text{bc}}_{\text{LB}}(\mathbf{x}, t)$ with help of $c_t$, $c_x$ following~\cref{eq:u_couple}.\\
                \STATE \hskip3.0em  9. Compute boundary density $\rho_{\text{bc}}(\mathbf{x}, t)$ thanks to~\cref{eq:density_bc}.\\
                \STATE \hskip3.0em  10. Reconstruct the particle's probability distribution function by combining~\cref{eq:f_i},\\ ~\cref{eq:f_i_neq} and \cref{eq:bc_eq}:\\
                \begin{equation}
                f^{\text{bc}}_i(\mathbf{x}, t)=\omega_{i}\rho_{\text{bc}}(\textbf{x},t) \bigg[1+\frac{\textbf{c}_i\cdot\mathbf{u}^{\text{bc}}_{\text{LB}}\left(\textbf{x},t\right)}{c_s^2}+\frac{\left[\textbf{c}_i\cdot\mathbf{u}^{\text{bc}}_{\text{LB}}\left(\textbf{x},t\right)\right]^2}{2c_s^4}  - \frac{\left[\mathbf{u}^{\text{bc}}_{\text{LB}}\left(\textbf{x},t\right)\cdot\mathbf{u}^{\text{bc}}_{\text{LB}}\left(\textbf{x},t\right)\right]}{2c_s^2}\bigg]+\frac{\omega_i}{c_s^4} \mathbf{Q}_i:\mathbf{\Pi}^{\text{bc}}_{\text{neq}}
                \end{equation}
                \STATE \hskip3.0em  11. Update the LES-LBM relaxation time $t_{\text{eff}}$ in~\cref{eq:tau_eff}.\\
            \hskip2.0em \textbf{end if}\\
            \STATE \hskip2.0em  12. Apply stream and collide to update the $f_i(\mathbf{x}, t)$ at each cell~\cref{eq:lbe}\\
        \hskip1.0em\textbf{end for}\\
    \textbf{end for}\\
   \end{algorithmic}
\end{algorithm}

Note that the proposed integration methodology is designed to be used at any RANS/LES-LBM interface for creating turbulence. The present work is a one-way coupling, which means that the RANS simulation has to perform first, then the LES-LBM simulation can start. Nonetheless, it is possible to dynamically couple the hybrid RANS/LES using FVM. One interesting work can be found in~\cite{carlsson2022investigation}. Also, the present work uses the $k-\omega$ RANS model and the Smagorinsky LES model, other RANS models and LES models~\cite{spalart1992one} can also follow the same logic presented in the above algorithm.
\begin{figure}[t]
\centering
\includegraphics[width=1.\textwidth]{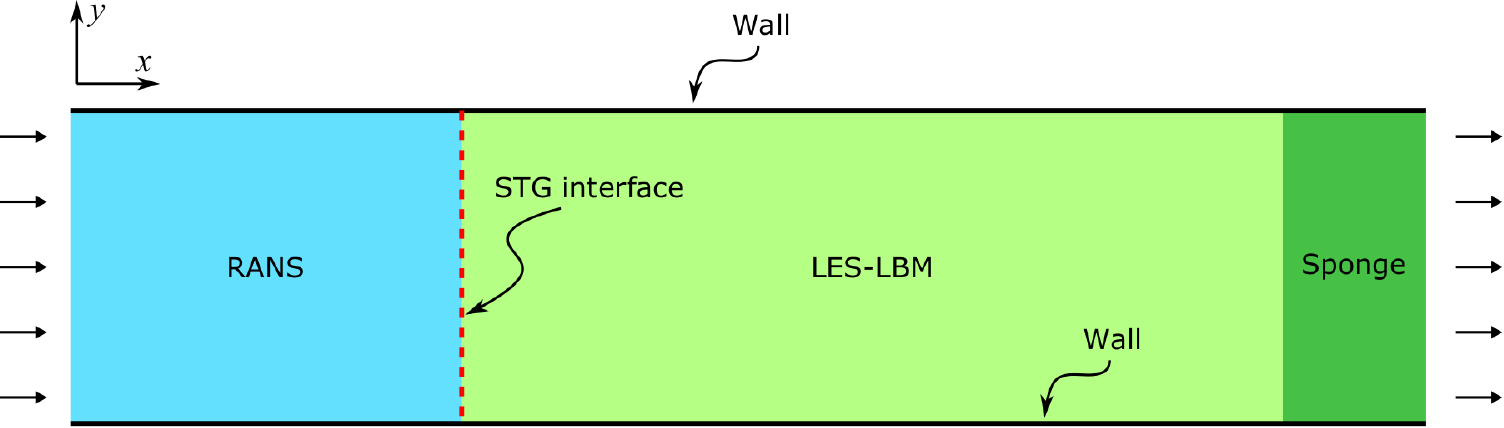}
\caption{2D sketch of the RANS/LES-LBM simulation. The blue part is the RANS simulation, and the green part is the LES-LBM simulation. The dark red dotted line is the RANS/LES-LBM interface which applies the STG to generate turbulent fluctuations as inflow fluctuations in the LES-LBM simulation. The outlet is imposed with the outflow pressure free boundary condition. The sponge layer is the dark green region adjacent to the outflow boundary.} 
\label{fig:sketch}
\end{figure}

\section{Numerical setup}\label{sec:set-up}
This section will present the channel flow simulation for the hybrid RANS/LES-LBM algorithm. The channel flow streamwise direction is the $x$ direction, the spanwise direction is the $z$ direction, and the wall-normal direction is the $y$ direction. Figure~\ref{fig:sketch} shows the 2D sketch for the hybrid RANS/LES-LBM channel flow simulation. The blue region represents the simulation with a finite-volume method~\cite{pyCALC-RANS}. 

In the RANS simulation, we employ the Wilcox $k-\omega$ RANS model~\cite{wilcox:88}. $y^+$ for the first wall cell centers is 0.6. Then the cell size is increased by $4\%$ toward the center of the channel where $\Delta y^+ \approx 8$. The total mesh in the $y$ direction has 96 cells. The extent of the domain is 2m. One cell is used in the $x$ and $z$ direction with homogeneous Neumann boundary conditions. The kinematic viscosity is set to $\nu_{\text{phys}} = 1/180 \mathrm{m^2/s}$. A constant volume force of $1\mathrm{N/m^3}$ is applied in the streamwise direction.

Following the STG-LBM algorithm, the RANS results will be set up as the synthetic turbulent generator inflow for the LES-LBM computational domain. The computational domain for the physical system is $L_x \times L_y \times L_z$, where $L_x = 19.2$m, $L_y = 2$m and $L_z=1.6$m. Let's set the height of the domain equal to $2\delta$, where $\delta=1$m. The red dotted line in Fig.~\ref{fig:sketch} represents the inlet STG boundary condition for the LBM. The green dotted line represents the regularized outlet pressure-free boundary condition. The no-slip bounce-back scheme is employed at the top and bottom walls in the $y$ direction. The periodic boundary condition is set for the spanwise direction. First, RANS simulation part will be computed with the friction Reynolds number $\mathrm{Re}_{\tau} = 180$. The bulk velocity at the RANS/LES-LBM interface is $U_{b} = 15.37\mathrm{m/s}$ in~\cref{eq:psu}. Also, the maximum velocity at the interface from the RANS simulation is $U_{\text{max}}=17.88\mathrm{m/s}$. For the LES-LBM part, to ensure the wall-resolved turbulent channel flow ($y^+ = 1$), we need at least 180 lattice cells in y direction. Therefore, the conversion factor of $c_x$ is set to $c_x=1/90$m. In this work, we use the uniform grid for the lattice Boltzmann simulation, thus the computational domain is set to $1728\times180\times144$ lattice Boltzmann units (LBU). The Mach number of the simulation can be defined as
\begin{equation}
\label{eq:ma-eq}
M_{a} = \frac{u_{\text{LB}}}{c_s},
\end{equation}
where $u_{\text{LB}}$ is the maximum mean flow in the LBM simulation. To ensure the stability of the BGK collision scheme of the LBM, the Mach number should be set to smaller than 0.1. We choose to set the BGK relaxation time equal to $\tau = 0.5025$ with the Smagorinsky constant equal to $C = 0.01$. From~\cref{eq:lbe-nu}, we can obtain the kinematic viscosity in the lattice Boltzmann simulation $\nu$. Combining~\cref{eq:c_factor}, the time conversion factor can be rewrote as $c_t = \nu/\nu_{phys}c_x^2$ which is $c_t = 1.85\times10^{-5}$s. Therefore, we can calculate $u_{\text{LB}}$ from $U_{\text{max}}$ by $u_{\text{LB}} = U_{\text{max}}c_t/c_x$. With help of~\cref{eq:ma-eq}, we obtain the Mach number which is equal to $M_{a}=0.0516$.
The dark region in~\cref{fig:sketch} is the sponge zone of the simulation, which is set to $0.4\delta$. The sponge zone is set up as a damping region to reduce the velocity fluctuations near the outlet~\cite{guo1994comparison, adams1998direct, fan2021source}. In the sponge zone, we impose a higher kinematic viscosity, $\nu_{\text{sponge}}$, as
\begin{equation}
\label{eq:sponge_visco}
\nu_{\text{sponge}} = \nu_{\text{eff}}\left [K\left (\frac{x-x_{\text{start}}}{x_{\text{end}}-x_{\text{start}}}\right)^{p}+1\right],
\end{equation}
where $K$ and $p$ are empirical constants, $\nu_{\text{eff}}$ is the effective viscosity calculated based on the Smagorinsky LES lattice Boltzmann model, $x_{\text{start}}$ and $x_{\text{end}}$ are the start point and end point of the sponge zone, respectively. The empirical constants are set to $K=1000$ and $p=3$ to ensure the stability of the simulation.
\begin{figure}[t]
\centering
\includegraphics[width=1\textwidth]{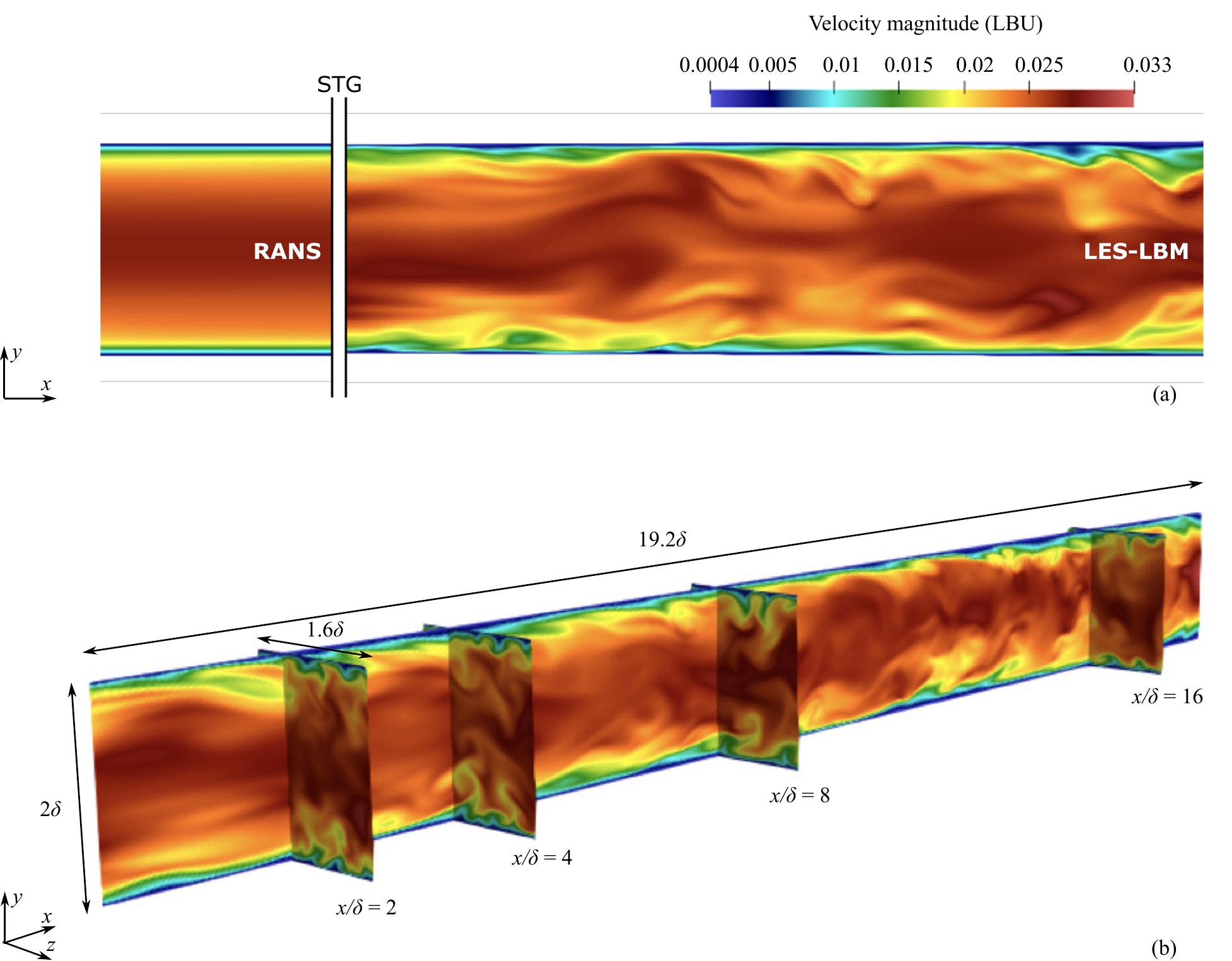}
\caption{Instantaneous velocity magnitude.  (a): the sketch of $xy$ plane. (b): Changing cross-sections along the streamwise direction for the LES-LBM simulation.} 
\label{fig:overview_channel}
\end{figure}

\section{Results}\label{sec:results}
This section presents the results of the channel flow simulation using the synthetic turbulent generator at the RANS/LES-LBM interface as inlet boundary condition. The dimensionless time unit for the channel flow simulation can be defined as the time that the bulk fluid passes the  boundary layer thickness, which can be written as
\begin{equation}
\label{eq:time_passflow}
t^* = \frac{\delta}{U_{\text{bulk}}},
\end{equation}
where $U_{\text{bulk}}$ is the bulk fluid velocity and $\delta$ is the half-channel height. 
The shear Reynolds number can be defined as
\begin{equation}
\label{eq:time_passflow}
\mathrm{Re}_{\tau} = \frac{u_{\tau}\delta}{\nu},
\end{equation}
where $u_{\tau}$ is the friction velocity and $\nu$ is the kinematic viscosity. The friction velocity is defined as  
\begin{equation}
\label{eq:u_shear}
u_{\tau} = \sqrt{\nu\frac{\partial{u_{\text{stream}}}}{\partial{y}}},
\end{equation}
where $u_{\text{stream}}$ is the ensemble average streamwise velocity at the first layer near the wall. The differential is calculated with the first-order approximation.
~\cref{fig:overview_channel}  shows the instantaneous velocity fields in the  $xy$ and $yz$ planes
The velocity profiles in \cref{fig:overview_channel}(b) are located at different cross-sections $x/\delta = 2, 4, 8, 16$. Qualitatively speaking, the flow downstream of the inlet quickly develops into turbulent flow downstream the inlet. The flow becomes less turbulent near the outlet due to the influence of the sponge layer. In the following part, we only focus on the LES-LBM part of the channel flow.

\begin{figure}[t]
\centering
\includegraphics[width=1\textwidth]{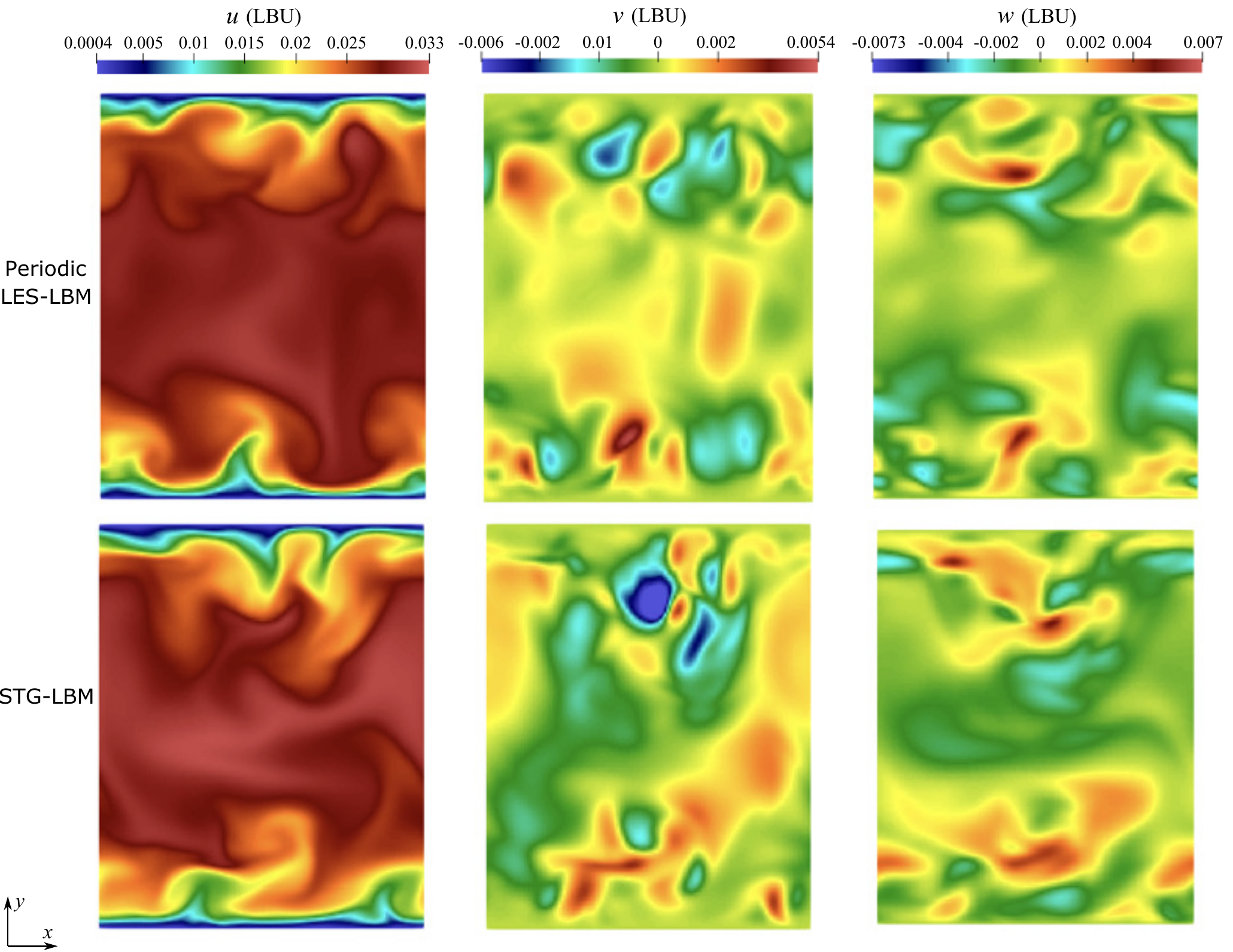}
\caption{Instantaneous 2D fields of $\mathrm{u}$, $\mathrm{v}$, $\mathrm{w}$ at the section of fully developed turbulent region.
The top panel represents the streamwise periodic case for the LES-LBM simulation results. The bottom panel represents the STG-generated LES-LBM results at the cross-section of $x = 9.6\delta$.}\label{fig:periodic-stg} 
\end{figure}
\subsection{Comparison with periodic boundary condition}
The initial conditions for the STG-LBM inlet-outlet simulations are $\mathrm{u}=\mathrm{v}=\mathrm{w}=0$, where $\mathrm{u}$, $\mathrm{v}$, $\mathrm{w}$ are representing velocity component in streamwise, spanwise and the vertical direction of the channel.  For comparison, we set up a periodic channel flow as a reference case (Periodic LES-LBM) by replacing the inlet/outlet boundary condition with the periodic boundary condition with the same $L_y = 2\delta$ and $L_z= 1.6\delta$. For the streamwise direction, we set $L_x=6.4\sigma$. The computational domain of the periodic LES-LBM is set to $576\times180\times144$ LBU with the uniform grid. Also, we keep the same parameters as the STG-LBM case for the Smagorinsky constant $C$, the relaxation time $\tau$, as well as conversion factors $c_x$ and $c_t$. We apply a volume force that is equal to $F=3.086\times 10^{-8}$ LBU. For the periodic case, the turbulence is triggered by placing a rectangular brick in the channel. 
After around 50 domain flow-through time, we removed the brick. We ran for another 130 domain flow-through time ($832t^{*}$) to fully develop turbulent flow.~\cref{fig:periodic-stg} shows a qualitative comparison between the streamwise periodic channel flow and the STG-LBM channel flow with inlet and outlet. The snapshot for STG is taken at  $x=8\delta$ at $t=58t^{*}$ from the abovementioned initial condition. Even if the STG method starts from a fairly poor initial condition (velocity field set to zero for the whole domain), the turbulence is well developed within a considerable short time. In contrast, for the periodic case, the time needing to develop the turbulent flow is highly dependent on the initial condition. Also, the periodic turbulent flow is limited to fundamental studies due to the "recycling" outflow.

Next, we will check the STG-LBM the adaptation length and the adaptation time, i.e. when the flow is fully developed in space and time. First, we will check the adaptation length by comparing the STG-LBM channel flow results with DNS, periodic LES-LBM and synthetic eddy method in LBM~\cite{fan2021source}.

\subsection{The adaptation length}\label{sec:5.3}
\begin{figure}[h]
\centering
\includegraphics[width=1\textwidth]{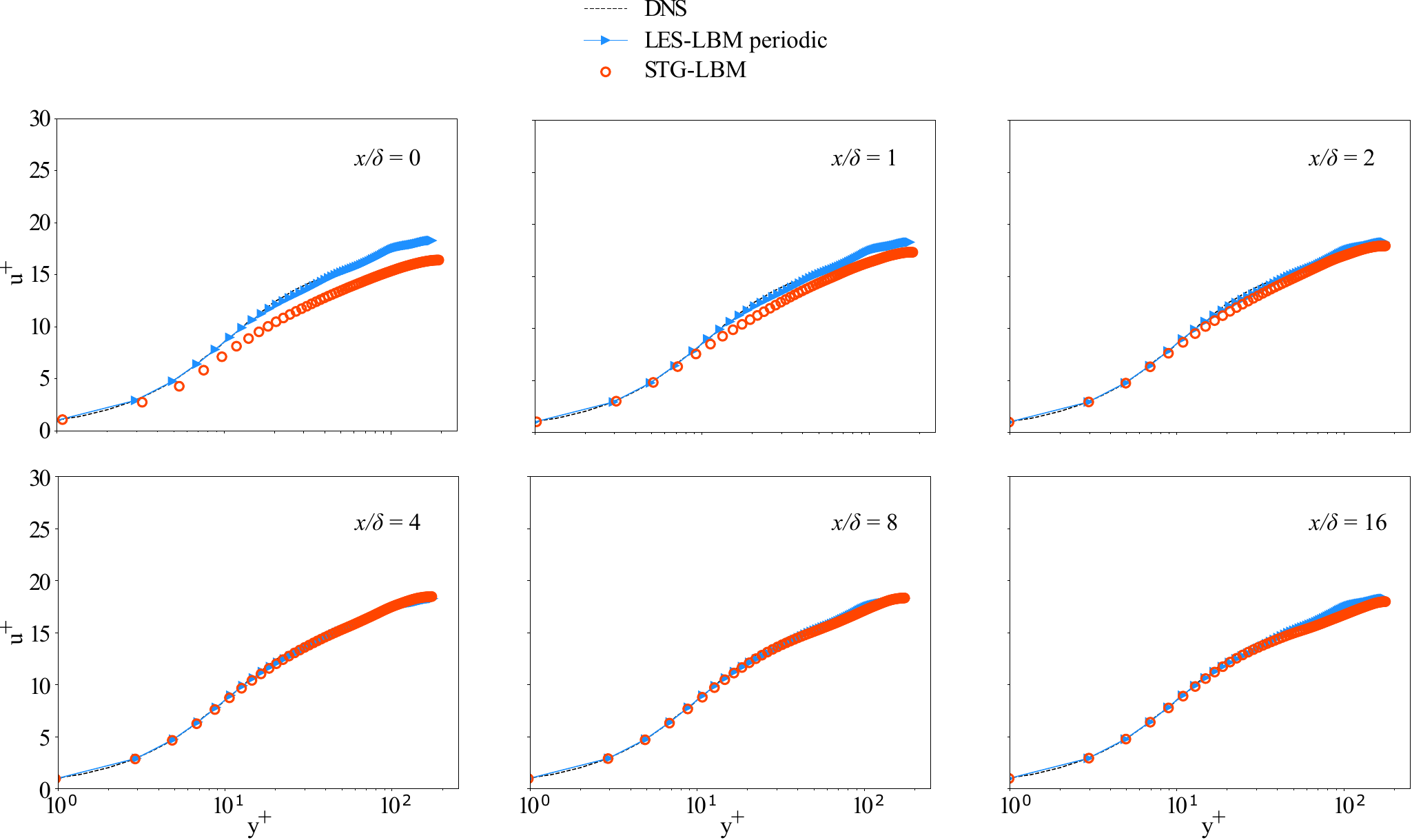}
\caption{$u^{+}$ as function of $y^{+}$. The black dotted lines are the DNS reference from Moser~\cite{moser1999direct}. The blue dots represent the LES-LBM periodic channel flow results.The red hollow round dots are the present STG-LBM LES results.} 
\label{fig:u_plus}
\end{figure}

In this subsection, all statistics for the STG-LBM channel flow simulations will be collected after around $2T$, where $T$ represents the time for the bulk velocity passing domain, which is $T=19.2t^*$. Then, all statistics will be gathered for another $8T$. ~\cref{fig:u_plus} shows $u^+$ as function of $y^+$ at different streamwise positions, where
\begin{equation}
\label{eq:y_plus}
 u^{+}=\frac{\left\langle u \right\rangle }{u_{\tau}},\hspace{.6in} y^{+}=y\frac{\sqrt{\tau_{w}/\rho}}{\nu},
\end{equation}
where $\tau_{w}$ is the wall shear stress.
In~\cref{fig:u_plus}, mean velocity profiles at different cross-section at the streamwise direction have been compared with the DNS data by Moser et al.~\cite{moser1999direct} and the periodic LES-LBM results. For $x/\delta < 2$, the STG-LBM captures the velocity near the wall well but gives slightly too small values for $y^{+}>10$. However, for $x/\delta \geq 2$ the STG-LBM predicts a  mean velocity field which agrees well with DNS data for all $y^{+}$. 

\begin{figure}[h]
\centering
\includegraphics[width=1\textwidth]{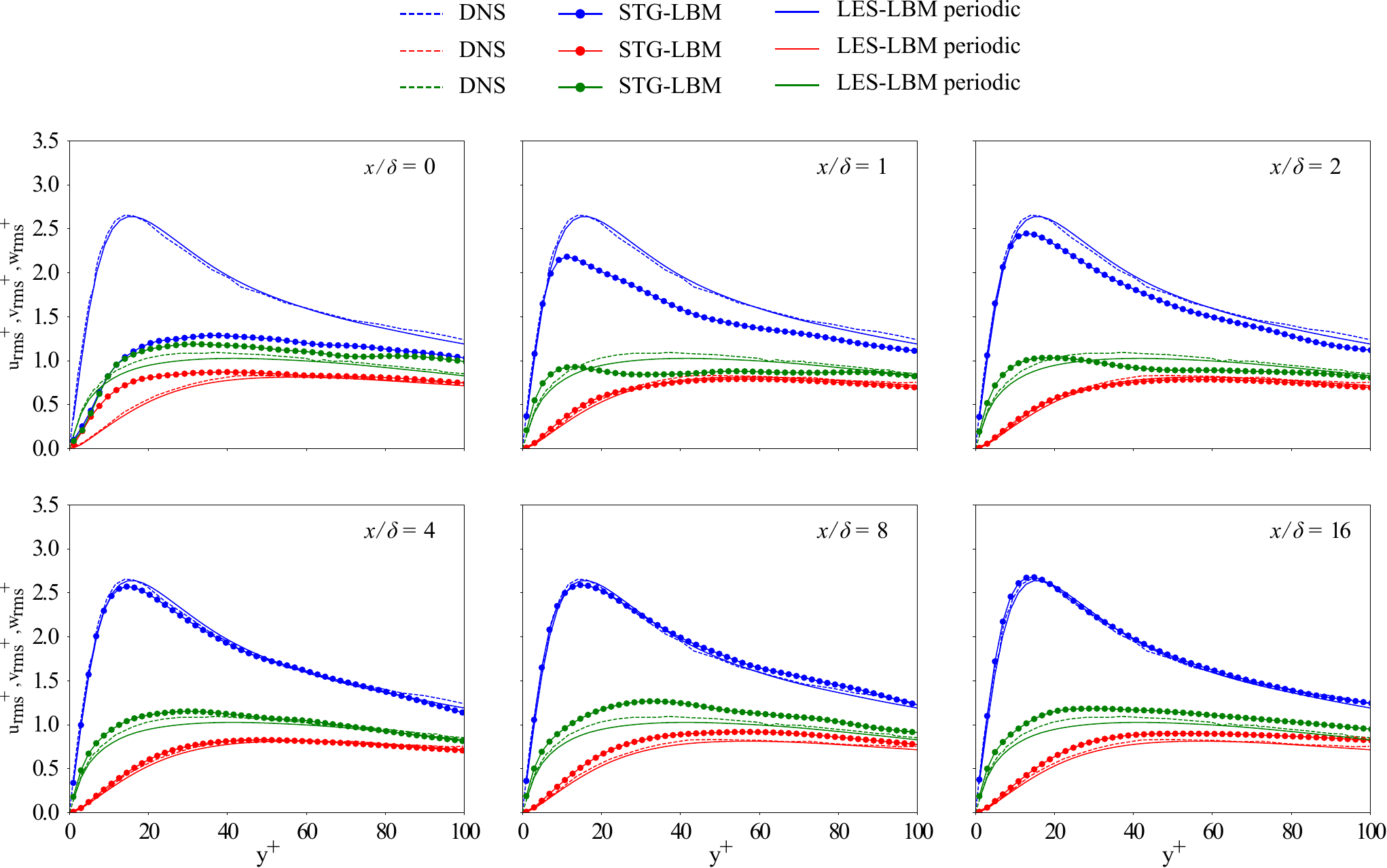}
\caption{$\mathrm{u}^{+}_{\text{rms}}$, $\mathrm{v}^{+}_{\text{rms}}$,$\mathrm{w}^{+}_{\text{rms}}$ as function of $y^{+}$ at different cross-sections along the streamwise direction. The dotted lines are the DNS reference from Moser~\cite{moser1999direct}. The round dots are the present STG-LBM LES results. The blue, red and green colors are representing $\mathrm{u}^{+}_{\text{rms}}$, $\mathrm{w}^{+}_{\text{rms}}$, $\mathrm{v}^{+}_{\text{rms}}$ respectively.}
\label{fig:RMS-space}
\end{figure}

Figure~\ref{fig:RMS-space} presents the $\mathrm{u}^{+}_{\text{rms}}$, $\mathrm{v}^{+}_{\text{rms}}$, $\mathrm{w}^{+}_{\text{rms}}$
as function of $y^{+}$ for DNS, periodic LBM and STG-LBM results. $\mathrm{u}^{+}_{\text{rms}}$, $\mathrm{v}^{+}_{\text{rms}}$, and  $\mathrm{w}^{+}_{\text{rms}}$ are the dimensionless root mean square(RMS) for three velocity components that are normalized with the shear velocity $u_{\tau}$:
\begin{equation}
\label{eq:v_rms}
\mathrm{u}^{+}_{\text{rms}} = \frac{\sqrt{\frac{\sum^{t_{\text{end}}}_{i=t_{\text{start}}}(u_i-\left\langle u \right\rangle)^2\Delta t}{\sum^{t_{\text{end}}}_{i=t_{\text{start}}}\Delta t}}}{u_{\tau}},
\end{equation}
where $\left\langle \cdot \right\rangle$ denotes 
the velocity field averaged in time.
Similar to~\cref{fig:u_plus}, STG-LBM failed to predict the normalized RMS results for the cross section at $x/\delta = 0$ and $x/\delta = 1$. This is reasonable since the inflow STG needs some space to develop into fully turbulent flow. For $x/\delta \ge 2$, the STG-LBM LES simulations agree well with DNS data and the periodic LES-LBM reference. 

\begin{figure}[h]
\centering
\includegraphics[width=1\textwidth]{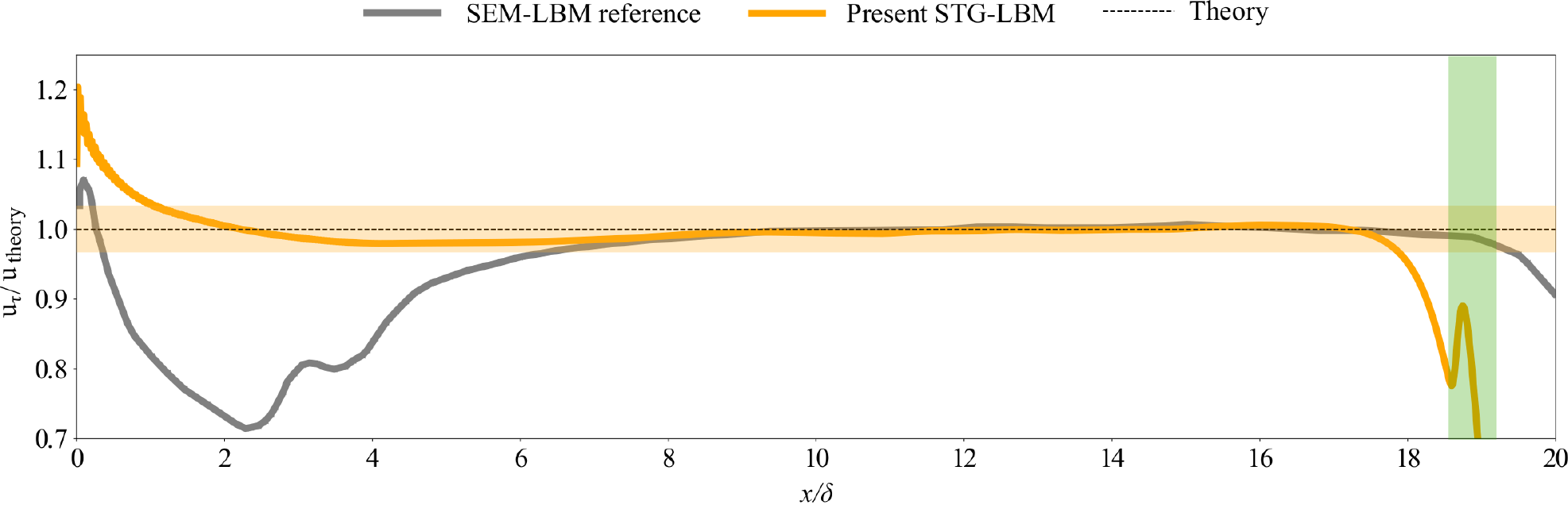}
\caption{Normalized $\mathrm{u}_{\tau}$ as function of $x/\delta$, the green region represents the sponge zone. The black dotted line is the theoretical data from $\mathrm{Re}_{\tau} = 180$. The yellow line is the STG-LBM results. The grey line is the SEM-LBM reference~\cite{fan2021source}.
}
\label{fig:u_tau_normed}
\end{figure}
Figure~\ref{fig:u_tau_normed} presents the normalized shear velocity $u_{\tau}/u_{\text{theory}}$ as function of $x/\delta$ along the channel. 
$u_{\text{theory}}$ is obtained from a force balance between the driving volume force (pressure gradient) and the wall shear stresses.
The present STG-LBM results have been compared with the SEM-LBM results~\cite{fan2021source}. The orange shadow is the trust region (TR) for the normalized shear velocity. It has an error bar of $\pm 3\%$. The green region is the sponge zone before the outlet boundary condition. For the STG-LBM, the channel flow simulation ends at $x/\delta = 19.2$ and for the SEM-LBM simulation, the channel flow simulation ends at $x/\delta = 20.0$. For the STG-LBM case, the normalized shear velocity has a peak near the inlet and enters the TR around $x/\delta = 1.7$. The results in ~\cref{fig:u_tau_normed}, ~\cref{fig:u_plus} and~\cref{fig:RMS-space} show that the STG-LBM has an adaptation length of around $x/\delta = 2$. Near the outlet (for  $x/\delta > 18.0$), the normalized shear decreases  due to the influence of the sponge zone. For SEM-LBM case, the SEM forcing takes place in the region of $x/\delta = 3.0 \pm 0.8$, and the normalized shear velocity becomes trustworthy after $x/\delta = 7.5$. As reported in~\cite{fan2021source}, the SEM-LBM is able to fully converge to the DNS data of the mean velocity profile and the Reynolds shear stress at $x/\delta = 18$. 
Next, we are interested to exam
the adaptation time, i.e. how long we can get the high-quality turbulent results. Notice that we were not able to find the information about the adaptation time in the lattice Boltzmann related turbulence generation method. However, it is crucial to show this quantity to potential readers who are interested to check how long it takes to generate the turbulent flows. According to the authors experience, it is non-trivial to achieve high-quality turbulent flow within short simulation time.
%
\begin{figure}[h]
\centering
\includegraphics[width=1\textwidth]{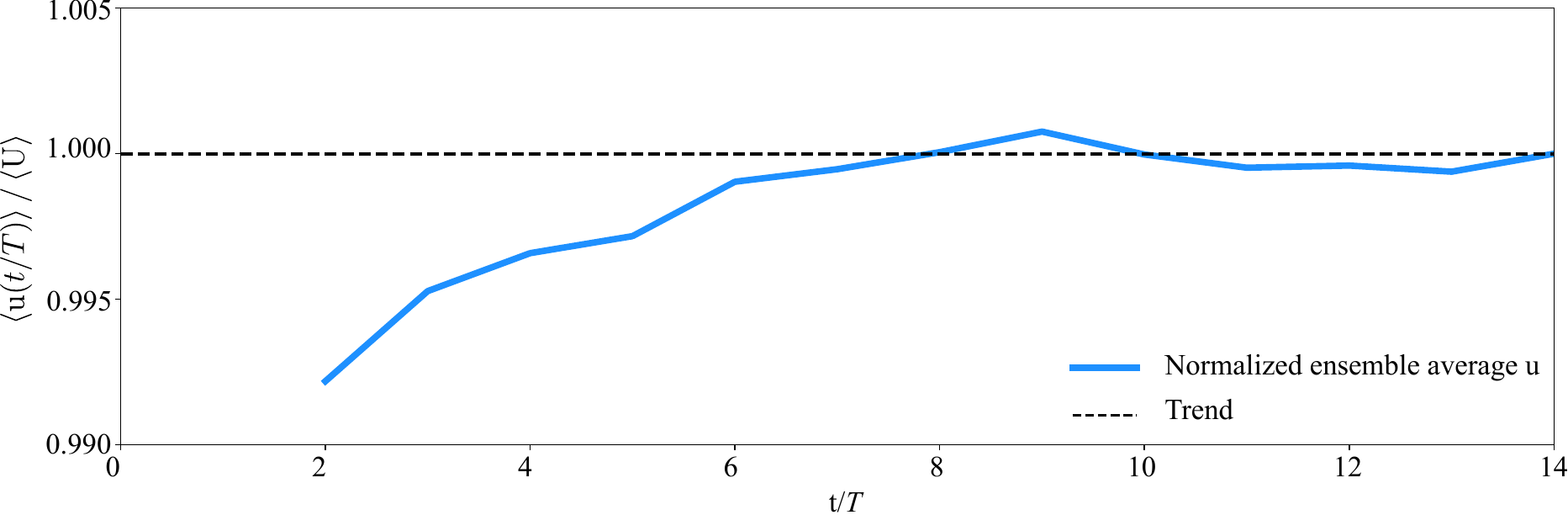}
\caption{Normalized ensemble average velocity, $\left \langle \mathrm{u}(t/T)\right \rangle/\left \langle \mathrm{U}\right \rangle$, as function of $t/T$ at the cross-section of $x/\delta=8$ with $y^+ = 100$. $\left \langle \mathrm{U}\right \rangle$ denotes the mean velocity at $y^+ = 100$. The statistics start from $1T$.}
\label{fig:tke-norm}
\end{figure}
\subsection{The adaptation time}
This subsection focuses on the time convergence study for turbulent channel flow using the proposed STG-LBM LES method.
Figure~\ref{fig:tke-norm} shows the normalized ensemble average velocity at the cross-section $x/\delta=8$ with $y^+=100$ as function of normalized time unit $t/T$. The ensemble average starts from $1T$ with interval of $1T$. For instance, first point at $2T$ in~\cref{fig:tke-norm} represents the ensemble average velocity from $1T$ to $2T$. We can observe that the ensemble average velocity converges to 1 within $1\%$ of the difference to the mean velocity, $\left \langle U \right \rangle$, at $y^+=100$. A distinct advantage is that the present work shows a good convergence rate even at $1T-2T$, demonstrating the good potential for a speedy converging method. We further check the adaptation time for both mean velocity profile and velocities RMS to with quantitative analysis by comparing with DNS data.

\begin{figure}[h]
\centering
\includegraphics[width=0.67\textwidth]{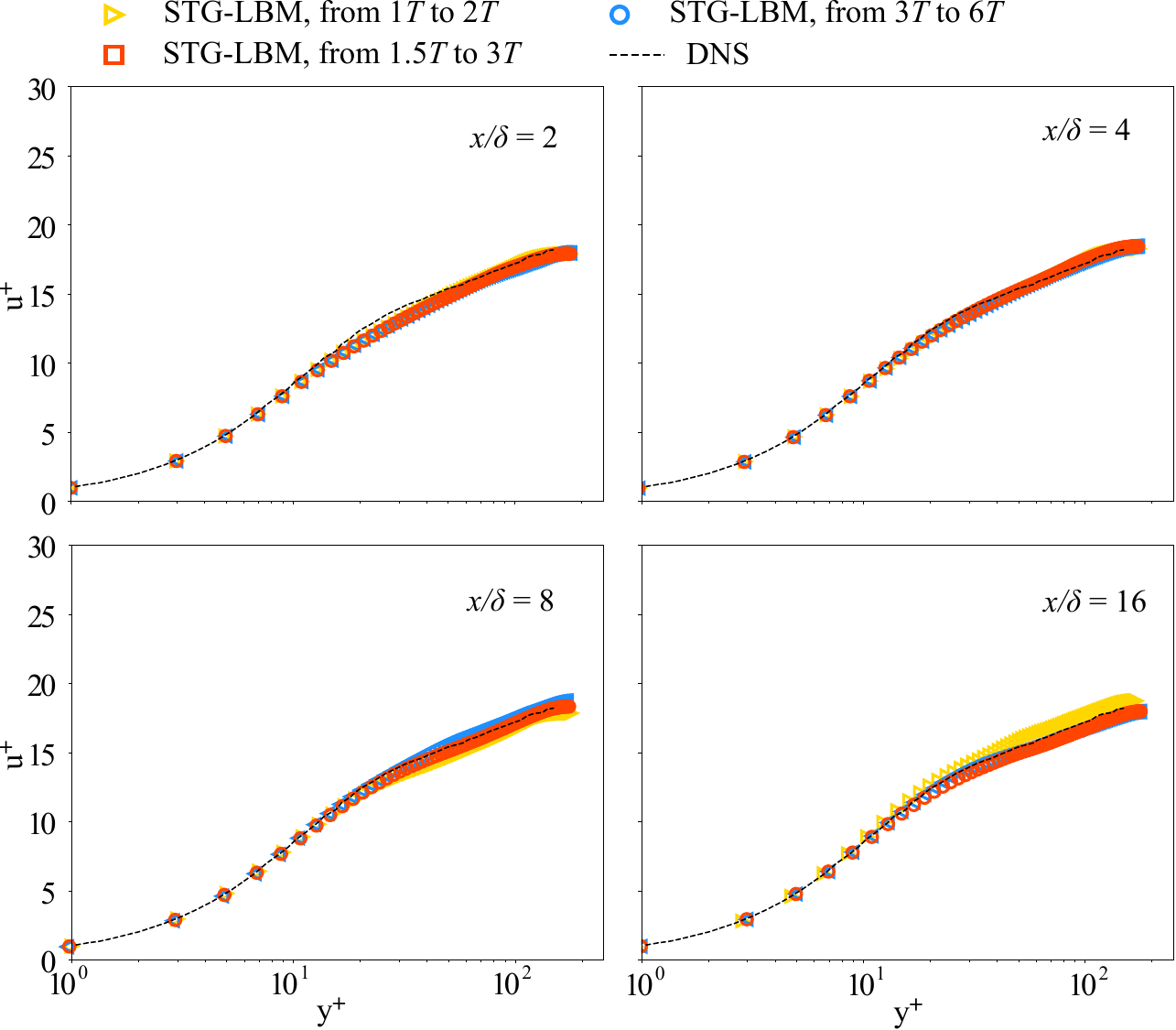}
\caption{$u^{+}$ as function of $y^{+}$ at changing $x/\delta$ and at different simulation time. Ensemble time and space average of $u^{+}$ at different cross-sections along the streamwise direction from $x/\delta = 2$ to $x/\delta = 16$.  The black dotted lines are the DNS reference from Moser~\cite{moser1999direct}. The yellow triangles, the blue round dots, and the red squares are the ensemble time average of $u^{+}$ from $1T$ to $2T$, from $1.5T$ to $3T$, and from $3T$ to $6T$, respectively.
}
\label{fig:time-u_plus}
\end{figure}

According to~\cref{fig:u_plus}, STG-LBM results match DNS data well for $x/\delta \ge 2$. In~\cref{fig:time-u_plus}, we present STG-LBM ensemble time and space average of $u^{+}$ at cross-sections along the streamwise direction from $x/\delta = 2$ to $x/\delta = 16$ at various time intervals, namely, from $1T$ to $2T$, from $1.5T$ to $3T$, from $3T$ to $6T$.
At $t=2T$, the STG-LBM shows acceptable discrepancies at $x/\delta=8$ and $x/\delta=16$ with DNS data. However, for $t=3T$ and $t=6T$ cases, both results almost overlap each other and show good agreement with the DNS data. Thus, we need to further check for the velocity RMS of these cases.

\begin{figure}[h]
\centering
\includegraphics[width=1\textwidth]{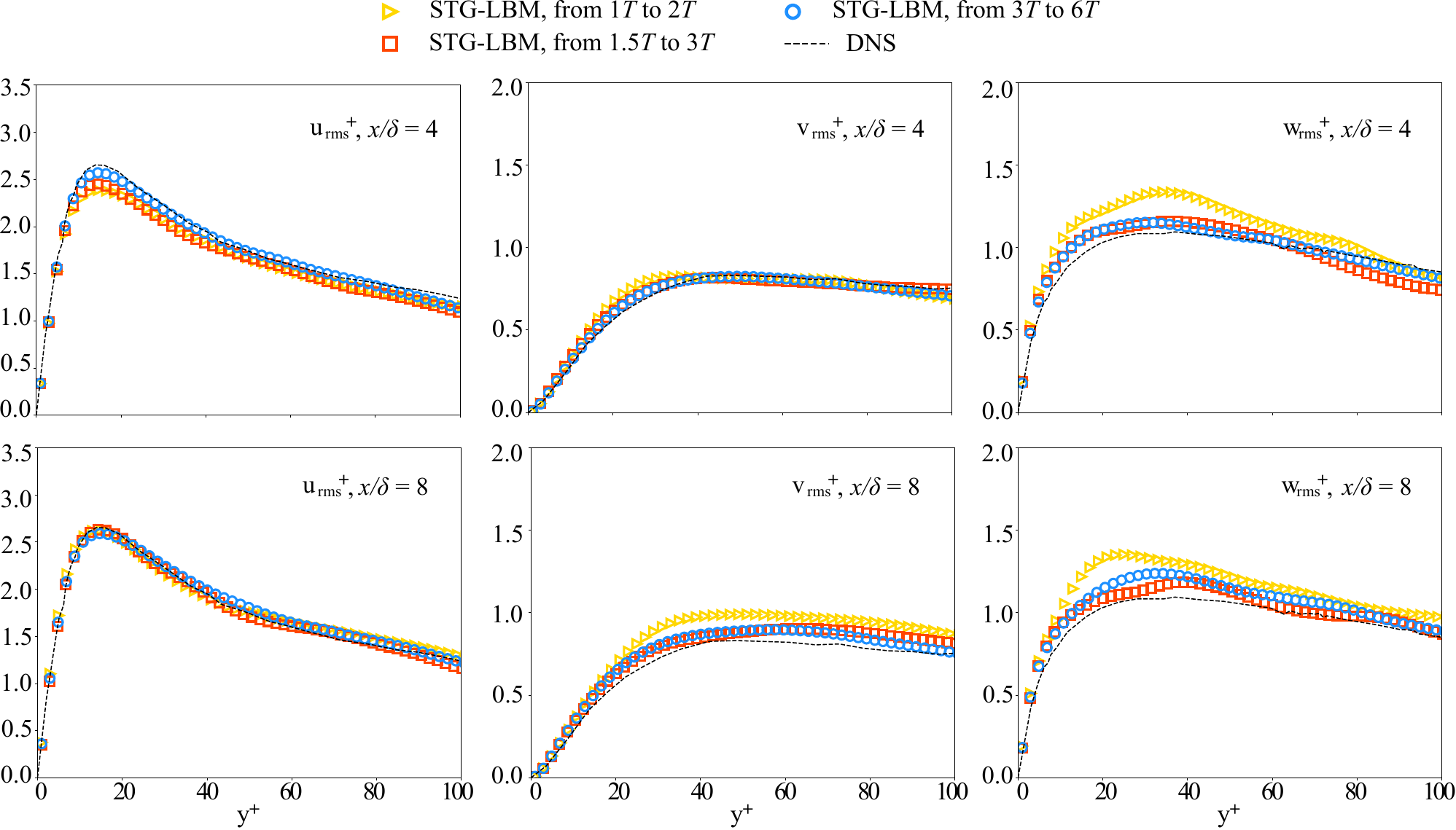}
\caption{$\mathrm{u}^{+}_{\text{rms}}$, $\mathrm{v}^{+}_{\text{rms}}$,$\mathrm{w}^{+}_{\text{rms}}$ as function of $y^{+}$ at different simulation time $T$. The dotted lines are the DNS reference from Moser~\cite{moser1999direct}. The round dots are the present STG-LBM LES results. The left, middle and right panels show results for $\mathrm{u}^{+}_{\text{rms}}$, $\mathrm{v}^{+}_{\text{rms}}$, and $\mathrm{w}^{+}_{\text{rms}}$. The yellow triangles, the blue round dots, and the red squares are the ensemble time average of RMS velocties from $1T$ to $2T$, from $1.5T$ to $3T$, and from $3T$ to $6T$, respectively.
}
\label{fig:time-RMS}
\end{figure}

Figure~\ref{fig:time-RMS} shows the RMS of the three velocity components $\mathrm{u}^{+}_{\text{rms}}$, $\mathrm{v}^{+}_{\text{rms}}$, $\mathrm{w}^{+}_{\text{rms}}$ as function of $y^{+}$ at cross section of $x/\delta = 4$ and $x/\delta = 8$. 
As can be seen, for $t=3T$ and $t=6T$, all stresses agree reasonably well with DNS data both at $x/\delta=4$  and $x/\delta=8$. For $t=2T$ case, larger discrepancies have been observed for $\mathrm{v}^{+}_{\text{rms}}$ and $\mathrm{w}^{+}_{\text{rms}}$ cases.

The quantitative results illustrate that the STG-LBM framework presents a surprisingly short adaptation time with around $t=1.5T$ to $t =3T$ which is $28.8 t^{*}$ to $57.6 t^{*}$. 
\subsection{Auto-correlation}
\begin{figure}[h]
\centering
\includegraphics[width=0.5\textwidth]{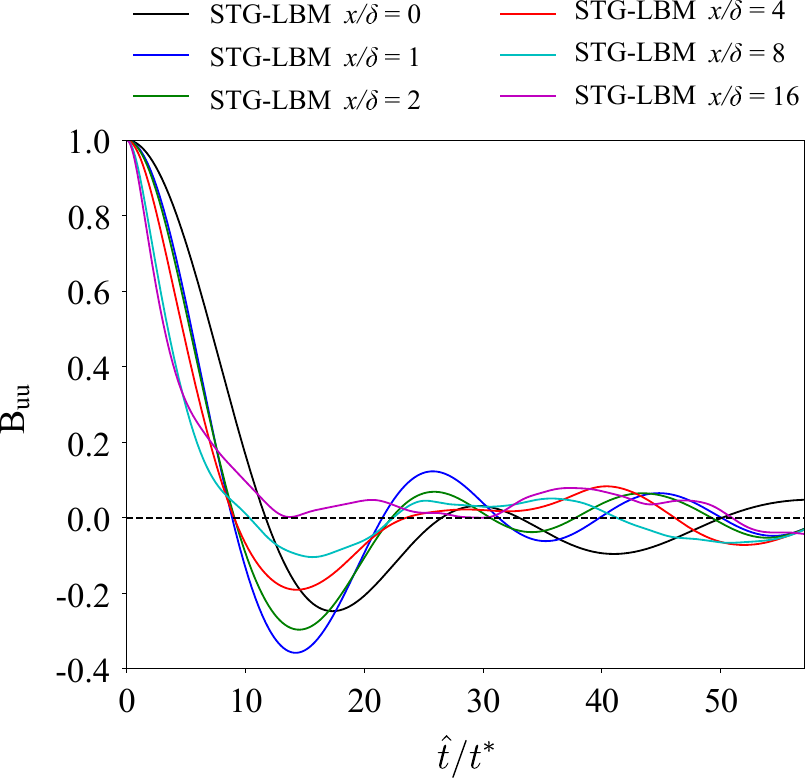}
\caption{The auto-correlation for the streamwise velocity, $\mathrm{B}_{\mathrm{uu}}$, as function of the normalized time separation $\hat{t}/t^*$ at different cross-sections $x/\delta = 0, 1, 2, 4, 8, 16$ with $y^{+} = 100$.}
\label{fig:auto-corr}
\end{figure}

Figure~\ref{fig:auto-corr} shows auto-correlation, $\mathrm{B}_{\mathrm{uu}}$, for the streamwise velocity as function of normalized time separation $\hat{t}/t^{*}$, here $\hat{t} = t - 3T$. We checked cells at different cross-sections $x/\delta = 0, 1, 2, 4, 8, 16$ at $y^{+} = 100$. The statistics starts to collect after $t=3T$. For all cases, the auto-correlation function drops quickly after the statistics start to be collected. Then, they reach a minimum at around $\hat{t}/t^{*}=14$ to $\hat{t}/t^{*}=16$. After that, all curves oscillate around 0. Similar effects have also been observed in the literature~\cite{favier2010space}. From~\cref{fig:auto-corr}, we can calculate the integral time length, $T_{t}$, as
\begin{equation}
\label{eq:time-integral}
T_{t} = \int_0^{t_{\text{min}}} \mathrm{B}_{\mathrm{uu}}(\hat{t}) d\hat{t},
\end{equation}
where $\hat{t}_{\text{min}}$ is the time separation when $\mathrm{B}_{\mathrm{uu}}$ reaches zero.~\cref{tab:time-integral} shows the integral time length.
$T_{t}$ decreases when $x/\delta$ reaches 8 and increases to 4.91 at $x/\delta = 16$.An increase at the last $x$ station is probably due to the sponge zone. Hence, we find that the integral time scale increases for increasing $x$ which indicates that we impose too large a time scale on the STG fluctuations.
\begin{table}[H]
  \centering
  \caption{The integral time length at different cross-sections at $y^{+} = 100$.}
    \begin{tabular}{cc}
    \hline
    \hline
    Cross section  & $T_{t}(\hat{t}/t^*)$ \\
    \hline
    $x/\delta = 0$ & $6.08\hat{t}/t^*$\\
    $x/\delta = 1$ & $4.53\hat{t}/t^*$ \\
    $x/\delta = 2$ & $3.87\hat{t}/t^*$ \\
    $x/\delta = 4$ & $3.65\hat{t}/t^*$ \\
    $x/\delta = 8$ & $2.67\hat{t}/t^*$ \\
    $x/\delta = 16$ & $4.91\hat{t}/t^*$ \\
    \hline
    \hline
    \end{tabular}%
  \label{tab:time-integral}%
\end{table}%

\section{Conclusions}\label{sec:conclusions}
This paper proposes a methodology to integrate the synthetic turbulent generator (STG) into the lattice Boltzmann framework. The proposed method can be used to couple the hybrid RANS/LES-LBM model with the STG interface. The present work uses the channel flow at $\mathrm{Re}_{\tau} = 180$ as a showcase, the RANS part is handled by a finite volume solver and the LES part is calculated via the LBM framework. At the RANS/LES-LBM interface, the STG serves as the inlet for the LBM channel flow. We qualitatively compare the STG-LBM channel flow results with those of a periodic LES-LBM case. We find good agreement on the turbulent structure. Compared with the SEM-LBM, periodic LES-LBM, and DNS channel flow data, the STG-LBM framework reaches close-to-DNS turbulent flow already at $x/\delta=2$. However, for the SEM-LBM case, it takes $x/\delta = 15$ to get similar results.
Moreover, we check the time convergence speed by comparing DNS data and recording the mean velocity and the velocity RMS at different time. Results show that the STG-LBM can reach a fully developed turbulent state around $t=1.5T$ to $t=3T$. The STG-LBM shows delightful potential in fluid/aero-dynamics-related applications, which requires fast convergence speed for fully developed turbulent conditions. 
\section{Acknowledgements}
The authors kindly acknowledge the funding from Chalmers Transport Area of Advance. The computations and data handling were enabled by resources provided by the Swedish National Infrastructure for Computing (SNIC), partially funded by the Swedish Research Council through grant agreement no. 2018-05973.

%
%

%


\end{CJK*}
\bibliography{prex}

\hyphenation{Post-Script Sprin-ger}
\begin{thebibliography}{52}%
\makeatletter
\providecommand \@ifxundefined [1]{%
 \@ifx{#1\undefined}
}%
\providecommand \@ifnum [1]{%
 \ifnum #1\expandafter \@firstoftwo
 \else \expandafter \@secondoftwo
 \fi
}%
\providecommand \@ifx [1]{%
 \ifx #1\expandafter \@firstoftwo
 \else \expandafter \@secondoftwo
 \fi
}%
\providecommand \natexlab [1]{#1}%
\providecommand \enquote  [1]{``#1''}%
\providecommand \bibnamefont  [1]{#1}%
\providecommand \bibfnamefont [1]{#1}%
\providecommand \citenamefont [1]{#1}%
\providecommand \href@noop [0]{\@secondoftwo}%
\providecommand \href [0]{\begingroup \@sanitize@url \@href}%
\providecommand \@href[1]{\@@startlink{#1}\@@href}%
\providecommand \@@href[1]{\endgroup#1\@@endlink}%
\providecommand \@sanitize@url [0]{\catcode `\\12\catcode `\$12\catcode
  `\&12\catcode `\#12\catcode `\^12\catcode `\_12\catcode `\%12\relax}%
\providecommand \@@startlink[1]{}%
\providecommand \@@endlink[0]{}%
\providecommand \url  [0]{\begingroup\@sanitize@url \@url }%
\providecommand \@url [1]{\endgroup\@href {#1}{\urlprefix }}%
\providecommand \urlprefix  [0]{URL }%
\providecommand \Eprint [0]{\href }%
\providecommand \doibase [0]{http://dx.doi.org/}%
\providecommand \selectlanguage [0]{\@gobble}%
\providecommand \bibinfo  [0]{\@secondoftwo}%
\providecommand \bibfield  [0]{\@secondoftwo}%
\providecommand \translation [1]{[#1]}%
\providecommand \BibitemOpen [0]{}%
\providecommand \bibitemStop [0]{}%
\providecommand \bibitemNoStop [0]{.\EOS\space}%
\providecommand \EOS [0]{\spacefactor3000\relax}%
\providecommand \BibitemShut  [1]{\csname bibitem#1\endcsname}%
\let\auto@bib@innerbib\@empty
\bibitem [{\citenamefont {Smagorinsky}(1963)}]{smagorinsky1963general}%
  \BibitemOpen
  \bibfield  {author} {\bibinfo {author} {\bibfnamefont {J.}~\bibnamefont
  {Smagorinsky}},\ }\bibfield  {title} {\enquote {\bibinfo {title} {General
  circulation experiments with the primitive equations: I. the basic
  experiment},}\ }\href@noop {} {\bibfield  {journal} {\bibinfo  {journal}
  {Monthly weather review}\ }\textbf {\bibinfo {volume} {91}},\ \bibinfo
  {pages} {99--164} (\bibinfo {year} {1963})}\BibitemShut {NoStop}%
\bibitem [{\citenamefont {Spalart}\ and\ \citenamefont
  {Allmaras}(1992)}]{spalart1992one}%
  \BibitemOpen
  \bibfield  {author} {\bibinfo {author} {\bibfnamefont {P.}~\bibnamefont
  {Spalart}}\ and\ \bibinfo {author} {\bibfnamefont {S.}~\bibnamefont
  {Allmaras}},\ }\bibfield  {title} {\enquote {\bibinfo {title} {A one-equation
  turbulence model for aerodynamic flows},}\ }in\ \href@noop {} {\emph
  {\bibinfo {booktitle} {30th aerospace sciences meeting and exhibit}}}\
  (\bibinfo {year} {1992})\ p.\ \bibinfo {pages} {439}\BibitemShut {NoStop}%
\bibitem [{\citenamefont {Menter}(1993)}]{menter1993zonal}%
  \BibitemOpen
  \bibfield  {author} {\bibinfo {author} {\bibfnamefont {F.}~\bibnamefont
  {Menter}},\ }\bibfield  {title} {\enquote {\bibinfo {title} {Zonal two
  equation kw turbulence models for aerodynamic flows},}\ }in\ \href@noop {}
  {\emph {\bibinfo {booktitle} {23rd fluid dynamics, plasmadynamics, and lasers
  conference}}}\ (\bibinfo {year} {1993})\ p.\ \bibinfo {pages}
  {2906}\BibitemShut {NoStop}%
\bibitem [{\citenamefont {Wu}(2017)}]{wu2017inflow}%
  \BibitemOpen
  \bibfield  {author} {\bibinfo {author} {\bibfnamefont {X.}~\bibnamefont
  {Wu}},\ }\bibfield  {title} {\enquote {\bibinfo {title} {Inflow turbulence
  generation methods},}\ }\href@noop {} {\bibfield  {journal} {\bibinfo
  {journal} {Annual Review of Fluid Mechanics}\ }\textbf {\bibinfo {volume}
  {49}},\ \bibinfo {pages} {23--49} (\bibinfo {year} {2017})}\BibitemShut
  {NoStop}%
\bibitem [{\citenamefont {Schl{\"u}ter}, \citenamefont {Pitsch},\ and\
  \citenamefont {Moin}(2004)}]{schluter2004large}%
  \BibitemOpen
  \bibfield  {author} {\bibinfo {author} {\bibfnamefont {J.}~\bibnamefont
  {Schl{\"u}ter}}, \bibinfo {author} {\bibfnamefont {H.}~\bibnamefont
  {Pitsch}}, \ and\ \bibinfo {author} {\bibfnamefont {P.}~\bibnamefont
  {Moin}},\ }\bibfield  {title} {\enquote {\bibinfo {title} {Large-eddy
  simulation inflow conditions for coupling with reynolds-averaged flow
  solvers},}\ }\href@noop {} {\bibfield  {journal} {\bibinfo  {journal} {AIAA
  journal}\ }\textbf {\bibinfo {volume} {42}},\ \bibinfo {pages} {478--484}
  (\bibinfo {year} {2004})}\BibitemShut {NoStop}%
\bibitem [{\citenamefont {Lund}, \citenamefont {Wu},\ and\ \citenamefont
  {Squires}(1998)}]{lund1998generation}%
  \BibitemOpen
  \bibfield  {author} {\bibinfo {author} {\bibfnamefont {T.~S.}\ \bibnamefont
  {Lund}}, \bibinfo {author} {\bibfnamefont {X.}~\bibnamefont {Wu}}, \ and\
  \bibinfo {author} {\bibfnamefont {K.~D.}\ \bibnamefont {Squires}},\
  }\bibfield  {title} {\enquote {\bibinfo {title} {Generation of turbulent
  inflow data for spatially-developing boundary layer simulations},}\
  }\href@noop {} {\bibfield  {journal} {\bibinfo  {journal} {Journal of
  computational physics}\ }\textbf {\bibinfo {volume} {140}},\ \bibinfo {pages}
  {233--258} (\bibinfo {year} {1998})}\BibitemShut {NoStop}%
\bibitem [{\citenamefont {Spalart}, \citenamefont {Strelets},\ and\
  \citenamefont {Travin}(2006)}]{spalart2006direct}%
  \BibitemOpen
  \bibfield  {author} {\bibinfo {author} {\bibfnamefont {P.}~\bibnamefont
  {Spalart}}, \bibinfo {author} {\bibfnamefont {M.}~\bibnamefont {Strelets}}, \
  and\ \bibinfo {author} {\bibfnamefont {A.}~\bibnamefont {Travin}},\
  }\bibfield  {title} {\enquote {\bibinfo {title} {Direct numerical simulation
  of large-eddy-break-up devices in a boundary layer},}\ }\href@noop {}
  {\bibfield  {journal} {\bibinfo  {journal} {International Journal of Heat and
  Fluid Flow}\ }\textbf {\bibinfo {volume} {27}},\ \bibinfo {pages} {902--910}
  (\bibinfo {year} {2006})}\BibitemShut {NoStop}%
\bibitem [{\citenamefont {Shur}\ \emph {et~al.}(2011)\citenamefont {Shur},
  \citenamefont {Spalart}, \citenamefont {Strelets},\ and\ \citenamefont
  {Travin}}]{shur2011rapid}%
  \BibitemOpen
  \bibfield  {author} {\bibinfo {author} {\bibfnamefont {M.}~\bibnamefont
  {Shur}}, \bibinfo {author} {\bibfnamefont {P.~R.}\ \bibnamefont {Spalart}},
  \bibinfo {author} {\bibfnamefont {M.}~\bibnamefont {Strelets}}, \ and\
  \bibinfo {author} {\bibfnamefont {A.}~\bibnamefont {Travin}},\ }\bibfield
  {title} {\enquote {\bibinfo {title} {A rapid and accurate switch from rans to
  les in boundary layers using an overlap region},}\ }\href@noop {} {\bibfield
  {journal} {\bibinfo  {journal} {Flow, turbulence and combustion}\ }\textbf
  {\bibinfo {volume} {86}},\ \bibinfo {pages} {179--206} (\bibinfo {year}
  {2011})}\BibitemShut {NoStop}%
\bibitem [{\citenamefont {Mathey}\ \emph {et~al.}(2006)\citenamefont {Mathey},
  \citenamefont {Cokljat}, \citenamefont {Bertoglio},\ and\ \citenamefont
  {Sergent}}]{mathey2006assessment}%
  \BibitemOpen
  \bibfield  {author} {\bibinfo {author} {\bibfnamefont {F.}~\bibnamefont
  {Mathey}}, \bibinfo {author} {\bibfnamefont {D.}~\bibnamefont {Cokljat}},
  \bibinfo {author} {\bibfnamefont {J.~P.}\ \bibnamefont {Bertoglio}}, \ and\
  \bibinfo {author} {\bibfnamefont {E.}~\bibnamefont {Sergent}},\ }\bibfield
  {title} {\enquote {\bibinfo {title} {Assessment of the vortex method for
  large eddy simulation inlet conditions},}\ }\href@noop {} {\bibfield
  {journal} {\bibinfo  {journal} {Progress in Computational Fluid Dynamics, An
  International Journal}\ }\textbf {\bibinfo {volume} {6}},\ \bibinfo {pages}
  {58--67} (\bibinfo {year} {2006})}\BibitemShut {NoStop}%
\bibitem [{\citenamefont {Jarrin}\ \emph {et~al.}(2009)\citenamefont {Jarrin},
  \citenamefont {Prosser}, \citenamefont {Uribe}, \citenamefont
  {Benhamadouche},\ and\ \citenamefont {Laurence}}]{jarrin2009reconstruction}%
  \BibitemOpen
  \bibfield  {author} {\bibinfo {author} {\bibfnamefont {N.}~\bibnamefont
  {Jarrin}}, \bibinfo {author} {\bibfnamefont {R.}~\bibnamefont {Prosser}},
  \bibinfo {author} {\bibfnamefont {J.-C.}\ \bibnamefont {Uribe}}, \bibinfo
  {author} {\bibfnamefont {S.}~\bibnamefont {Benhamadouche}}, \ and\ \bibinfo
  {author} {\bibfnamefont {D.}~\bibnamefont {Laurence}},\ }\bibfield  {title}
  {\enquote {\bibinfo {title} {Reconstruction of turbulent fluctuations for
  hybrid rans/les simulations using a synthetic-eddy method},}\ }\href@noop {}
  {\bibfield  {journal} {\bibinfo  {journal} {International Journal of Heat and
  Fluid Flow}\ }\textbf {\bibinfo {volume} {30}},\ \bibinfo {pages} {435--442}
  (\bibinfo {year} {2009})}\BibitemShut {NoStop}%
\bibitem [{\citenamefont {Skillen}, \citenamefont {Revell},\ and\ \citenamefont
  {Craft}(2016)}]{skillen2016accuracy}%
  \BibitemOpen
  \bibfield  {author} {\bibinfo {author} {\bibfnamefont {A.}~\bibnamefont
  {Skillen}}, \bibinfo {author} {\bibfnamefont {A.}~\bibnamefont {Revell}}, \
  and\ \bibinfo {author} {\bibfnamefont {T.}~\bibnamefont {Craft}},\ }\bibfield
   {title} {\enquote {\bibinfo {title} {Accuracy and efficiency improvements in
  synthetic eddy methods.}}\ }\href@noop {} {\bibfield  {journal} {\bibinfo
  {journal} {International Journal of Heat and Fluid Flow}\ }\textbf {\bibinfo
  {volume} {62}},\ \bibinfo {pages} {386--394} (\bibinfo {year}
  {2016})}\BibitemShut {NoStop}%
\bibitem [{\citenamefont {Roidl}, \citenamefont {Meinke},\ and\ \citenamefont
  {Schr{\"o}der}(2011)}]{roidl2011zonal}%
  \BibitemOpen
  \bibfield  {author} {\bibinfo {author} {\bibfnamefont {B.}~\bibnamefont
  {Roidl}}, \bibinfo {author} {\bibfnamefont {M.}~\bibnamefont {Meinke}}, \
  and\ \bibinfo {author} {\bibfnamefont {W.}~\bibnamefont {Schr{\"o}der}},\
  }\bibfield  {title} {\enquote {\bibinfo {title} {Zonal rans-les computation
  of transonic airfoil flow},}\ }in\ \href@noop {} {\emph {\bibinfo {booktitle}
  {29th AIAA Applied Aerodynamics Conference}}}\ (\bibinfo {year} {2011})\ p.\
  \bibinfo {pages} {3974}\BibitemShut {NoStop}%
\bibitem [{\citenamefont {Roidl}, \citenamefont {Meinke},\ and\ \citenamefont
  {Schr{\"o}der}(2012)}]{roidl2012zonal}%
  \BibitemOpen
  \bibfield  {author} {\bibinfo {author} {\bibfnamefont {B.}~\bibnamefont
  {Roidl}}, \bibinfo {author} {\bibfnamefont {M.}~\bibnamefont {Meinke}}, \
  and\ \bibinfo {author} {\bibfnamefont {W.}~\bibnamefont {Schr{\"o}der}},\
  }\bibfield  {title} {\enquote {\bibinfo {title} {A zonal rans--les method for
  compressible flows},}\ }\href@noop {} {\bibfield  {journal} {\bibinfo
  {journal} {Computers \& Fluids}\ }\textbf {\bibinfo {volume} {67}},\ \bibinfo
  {pages} {1--15} (\bibinfo {year} {2012})}\BibitemShut {NoStop}%
\bibitem [{\citenamefont {Klein}, \citenamefont {Sadiki},\ and\ \citenamefont
  {Janicka}(2003)}]{klein2003digital}%
  \BibitemOpen
  \bibfield  {author} {\bibinfo {author} {\bibfnamefont {M.}~\bibnamefont
  {Klein}}, \bibinfo {author} {\bibfnamefont {A.}~\bibnamefont {Sadiki}}, \
  and\ \bibinfo {author} {\bibfnamefont {J.}~\bibnamefont {Janicka}},\
  }\bibfield  {title} {\enquote {\bibinfo {title} {A digital filter based
  generation of inflow data for spatially developing direct numerical or large
  eddy simulations},}\ }\href@noop {} {\bibfield  {journal} {\bibinfo
  {journal} {Journal of computational Physics}\ }\textbf {\bibinfo {volume}
  {186}},\ \bibinfo {pages} {652--665} (\bibinfo {year} {2003})}\BibitemShut
  {NoStop}%
\bibitem [{\citenamefont {Di~Mare}\ \emph {et~al.}(2006)\citenamefont
  {Di~Mare}, \citenamefont {Klein}, \citenamefont {Jones},\ and\ \citenamefont
  {Janicka}}]{di2006synthetic}%
  \BibitemOpen
  \bibfield  {author} {\bibinfo {author} {\bibfnamefont {L.}~\bibnamefont
  {Di~Mare}}, \bibinfo {author} {\bibfnamefont {M.}~\bibnamefont {Klein}},
  \bibinfo {author} {\bibfnamefont {W.}~\bibnamefont {Jones}}, \ and\ \bibinfo
  {author} {\bibfnamefont {J.}~\bibnamefont {Janicka}},\ }\bibfield  {title}
  {\enquote {\bibinfo {title} {Synthetic turbulence inflow conditions for
  large-eddy simulation},}\ }\href@noop {} {\bibfield  {journal} {\bibinfo
  {journal} {Physics of Fluids}\ }\textbf {\bibinfo {volume} {18}},\ \bibinfo
  {pages} {025107} (\bibinfo {year} {2006})}\BibitemShut {NoStop}%
\bibitem [{\citenamefont {Davidson}(2007)}]{davidson2007using}%
  \BibitemOpen
  \bibfield  {author} {\bibinfo {author} {\bibfnamefont {L.}~\bibnamefont
  {Davidson}},\ }\bibfield  {title} {\enquote {\bibinfo {title} {Using
  isotropic synthetic fluctuations as inlet boundary conditions for unsteady
  simulations},}\ }in\ \href@noop {} {\emph {\bibinfo {booktitle} {Advances and
  applications in Fluid Mechanics}}}\ (\bibinfo {organization} {Citeseer},\
  \bibinfo {year} {2007})\BibitemShut {NoStop}%
\bibitem [{\citenamefont {Xie}\ and\ \citenamefont
  {Castro}(2008)}]{xie2008efficient}%
  \BibitemOpen
  \bibfield  {author} {\bibinfo {author} {\bibfnamefont {Z.-T.}\ \bibnamefont
  {Xie}}\ and\ \bibinfo {author} {\bibfnamefont {I.~P.}\ \bibnamefont
  {Castro}},\ }\bibfield  {title} {\enquote {\bibinfo {title} {Efficient
  generation of inflow conditions for large eddy simulation of street-scale
  flows},}\ }\href@noop {} {\bibfield  {journal} {\bibinfo  {journal} {Flow,
  turbulence and combustion}\ }\textbf {\bibinfo {volume} {81}},\ \bibinfo
  {pages} {449--470} (\bibinfo {year} {2008})}\BibitemShut {NoStop}%
\bibitem [{\citenamefont {Huang}, \citenamefont {Li},\ and\ \citenamefont
  {Wu}(2010)}]{huang2010general}%
  \BibitemOpen
  \bibfield  {author} {\bibinfo {author} {\bibfnamefont {S.}~\bibnamefont
  {Huang}}, \bibinfo {author} {\bibfnamefont {Q.}~\bibnamefont {Li}}, \ and\
  \bibinfo {author} {\bibfnamefont {J.}~\bibnamefont {Wu}},\ }\bibfield
  {title} {\enquote {\bibinfo {title} {A general inflow turbulence generator
  for large eddy simulation},}\ }\href@noop {} {\bibfield  {journal} {\bibinfo
  {journal} {Journal of Wind Engineering and Industrial Aerodynamics}\ }\textbf
  {\bibinfo {volume} {98}},\ \bibinfo {pages} {600--617} (\bibinfo {year}
  {2010})}\BibitemShut {NoStop}%
\bibitem [{\citenamefont {Shur}\ \emph {et~al.}(2014)\citenamefont {Shur},
  \citenamefont {Spalart}, \citenamefont {Strelets},\ and\ \citenamefont
  {Travin}}]{shur2014synthetic}%
  \BibitemOpen
  \bibfield  {author} {\bibinfo {author} {\bibfnamefont {M.~L.}\ \bibnamefont
  {Shur}}, \bibinfo {author} {\bibfnamefont {P.~R.}\ \bibnamefont {Spalart}},
  \bibinfo {author} {\bibfnamefont {M.~K.}\ \bibnamefont {Strelets}}, \ and\
  \bibinfo {author} {\bibfnamefont {A.~K.}\ \bibnamefont {Travin}},\ }\bibfield
   {title} {\enquote {\bibinfo {title} {Synthetic turbulence generators for
  rans-les interfaces in zonal simulations of aerodynamic and aeroacoustic
  problems},}\ }\href@noop {} {\bibfield  {journal} {\bibinfo  {journal} {Flow,
  turbulence and combustion}\ }\textbf {\bibinfo {volume} {93}},\ \bibinfo
  {pages} {63--92} (\bibinfo {year} {2014})}\BibitemShut {NoStop}%
\bibitem [{\citenamefont {Succi}(2001)}]{succi2001lattice}%
  \BibitemOpen
  \bibfield  {author} {\bibinfo {author} {\bibfnamefont {S.}~\bibnamefont
  {Succi}},\ }\href@noop {} {\emph {\bibinfo {title} {The lattice Boltzmann
  equation: for fluid dynamics and beyond}}}\ (\bibinfo  {publisher} {Oxford
  university press},\ \bibinfo {year} {2001})\BibitemShut {NoStop}%
\bibitem [{\citenamefont {Kr{\"u}ger}\ \emph {et~al.}(2017)\citenamefont
  {Kr{\"u}ger}, \citenamefont {Kusumaatmaja}, \citenamefont {Kuzmin},
  \citenamefont {Shardt}, \citenamefont {Silva},\ and\ \citenamefont
  {Viggen}}]{kruger2017lattice}%
  \BibitemOpen
  \bibfield  {author} {\bibinfo {author} {\bibfnamefont {T.}~\bibnamefont
  {Kr{\"u}ger}}, \bibinfo {author} {\bibfnamefont {H.}~\bibnamefont
  {Kusumaatmaja}}, \bibinfo {author} {\bibfnamefont {A.}~\bibnamefont
  {Kuzmin}}, \bibinfo {author} {\bibfnamefont {O.}~\bibnamefont {Shardt}},
  \bibinfo {author} {\bibfnamefont {G.}~\bibnamefont {Silva}}, \ and\ \bibinfo
  {author} {\bibfnamefont {E.~M.}\ \bibnamefont {Viggen}},\ }\bibfield  {title}
  {\enquote {\bibinfo {title} {The lattice boltzmann method},}\ }\href@noop {}
  {\bibfield  {journal} {\bibinfo  {journal} {Springer International
  Publishing}\ }\textbf {\bibinfo {volume} {10}},\ \bibinfo {pages} {978--3}
  (\bibinfo {year} {2017})}\BibitemShut {NoStop}%
\bibitem [{\citenamefont {Mari{\'e}}, \citenamefont {Ricot},\ and\
  \citenamefont {Sagaut}(2009)}]{marie2009comparison}%
  \BibitemOpen
  \bibfield  {author} {\bibinfo {author} {\bibfnamefont {S.}~\bibnamefont
  {Mari{\'e}}}, \bibinfo {author} {\bibfnamefont {D.}~\bibnamefont {Ricot}}, \
  and\ \bibinfo {author} {\bibfnamefont {P.}~\bibnamefont {Sagaut}},\
  }\bibfield  {title} {\enquote {\bibinfo {title} {Comparison between lattice
  boltzmann method and navier--stokes high order schemes for computational
  aeroacoustics},}\ }\href@noop {} {\bibfield  {journal} {\bibinfo  {journal}
  {Journal of Computational Physics}\ }\textbf {\bibinfo {volume} {228}},\
  \bibinfo {pages} {1056--1070} (\bibinfo {year} {2009})}\BibitemShut {NoStop}%
\bibitem [{\citenamefont {Belardinelli}\ \emph {et~al.}(2015)\citenamefont
  {Belardinelli}, \citenamefont {Sbragaglia}, \citenamefont {Biferale},
  \citenamefont {Gross},\ and\ \citenamefont {Varnik}}]{Belardinelli15}%
  \BibitemOpen
  \bibfield  {author} {\bibinfo {author} {\bibfnamefont {D.}~\bibnamefont
  {Belardinelli}}, \bibinfo {author} {\bibfnamefont {M.}~\bibnamefont
  {Sbragaglia}}, \bibinfo {author} {\bibfnamefont {L.}~\bibnamefont
  {Biferale}}, \bibinfo {author} {\bibfnamefont {M.}~\bibnamefont {Gross}}, \
  and\ \bibinfo {author} {\bibfnamefont {F.}~\bibnamefont {Varnik}},\
  }\bibfield  {title} {\enquote {\bibinfo {title} {Fluctuating multicomponent
  lattice boltzmann model},}\ }\href@noop {} {\bibfield  {journal} {\bibinfo
  {journal} {Phys. Rev. E}\ }\textbf {\bibinfo {volume} {91}},\ \bibinfo
  {pages} {023313} (\bibinfo {year} {2015})}\BibitemShut {NoStop}%
\bibitem [{\citenamefont {Xue}\ \emph {et~al.}(2020)\citenamefont {Xue},
  \citenamefont {Biferale}, \citenamefont {Sbragaglia},\ and\ \citenamefont
  {Toschi}}]{xue2020brownian}%
  \BibitemOpen
  \bibfield  {author} {\bibinfo {author} {\bibfnamefont {X.}~\bibnamefont
  {Xue}}, \bibinfo {author} {\bibfnamefont {L.}~\bibnamefont {Biferale}},
  \bibinfo {author} {\bibfnamefont {M.}~\bibnamefont {Sbragaglia}}, \ and\
  \bibinfo {author} {\bibfnamefont {F.}~\bibnamefont {Toschi}},\ }\bibfield
  {title} {\enquote {\bibinfo {title} {A lattice boltzmann study on brownian
  diffusion and friction of a particle in a confined multicomponent fluid},}\
  }\href@noop {} {\bibfield  {journal} {\bibinfo  {journal} {Journal of
  Computational Science}\ }\textbf {\bibinfo {volume} {47}},\ \bibinfo {pages}
  {101113} (\bibinfo {year} {2020})}\BibitemShut {NoStop}%
\bibitem [{\citenamefont {Xue}\ \emph {et~al.}(2018)\citenamefont {Xue},
  \citenamefont {Sbragaglia}, \citenamefont {Biferale},\ and\ \citenamefont
  {Toschi}}]{xue2018effects}%
  \BibitemOpen
  \bibfield  {author} {\bibinfo {author} {\bibfnamefont {X.}~\bibnamefont
  {Xue}}, \bibinfo {author} {\bibfnamefont {M.}~\bibnamefont {Sbragaglia}},
  \bibinfo {author} {\bibfnamefont {L.}~\bibnamefont {Biferale}}, \ and\
  \bibinfo {author} {\bibfnamefont {F.}~\bibnamefont {Toschi}},\ }\bibfield
  {title} {\enquote {\bibinfo {title} {Effects of thermal fluctuations in the
  fragmentation of a nanoligament},}\ }\href@noop {} {\bibfield  {journal}
  {\bibinfo  {journal} {Physical Review E}\ }\textbf {\bibinfo {volume} {98}},\
  \bibinfo {pages} {012802} (\bibinfo {year} {2018})}\BibitemShut {NoStop}%
\bibitem [{\citenamefont {Hou}\ \emph {et~al.}(1994)\citenamefont {Hou},
  \citenamefont {Sterling}, \citenamefont {Chen},\ and\ \citenamefont
  {Doolen}}]{hou1994lattice}%
  \BibitemOpen
  \bibfield  {author} {\bibinfo {author} {\bibfnamefont {S.}~\bibnamefont
  {Hou}}, \bibinfo {author} {\bibfnamefont {J.}~\bibnamefont {Sterling}},
  \bibinfo {author} {\bibfnamefont {S.}~\bibnamefont {Chen}}, \ and\ \bibinfo
  {author} {\bibfnamefont {G.}~\bibnamefont {Doolen}},\ }\bibfield  {title}
  {\enquote {\bibinfo {title} {A lattice boltzmann subgrid model for high
  reynolds number flows},}\ }\href@noop {} {\bibfield  {journal} {\bibinfo
  {journal} {arXiv preprint comp-gas/9401004}\ } (\bibinfo {year}
  {1994})}\BibitemShut {NoStop}%
\bibitem [{\citenamefont {Toschi}\ and\ \citenamefont
  {Bodenschatz}(2009)}]{toschi2009lagrangian}%
  \BibitemOpen
  \bibfield  {author} {\bibinfo {author} {\bibfnamefont {F.}~\bibnamefont
  {Toschi}}\ and\ \bibinfo {author} {\bibfnamefont {E.}~\bibnamefont
  {Bodenschatz}},\ }\bibfield  {title} {\enquote {\bibinfo {title} {Lagrangian
  properties of particles in turbulence},}\ }\href@noop {} {\bibfield
  {journal} {\bibinfo  {journal} {Annual review of fluid mechanics}\ }\textbf
  {\bibinfo {volume} {41}},\ \bibinfo {pages} {375--404} (\bibinfo {year}
  {2009})}\BibitemShut {NoStop}%
\bibitem [{\citenamefont {Karlin}, \citenamefont {Ferrante},\ and\
  \citenamefont {{\"O}ttinger}(1999)}]{karlin1999perfect}%
  \BibitemOpen
  \bibfield  {author} {\bibinfo {author} {\bibfnamefont {I.~V.}\ \bibnamefont
  {Karlin}}, \bibinfo {author} {\bibfnamefont {A.}~\bibnamefont {Ferrante}}, \
  and\ \bibinfo {author} {\bibfnamefont {H.~C.}\ \bibnamefont {{\"O}ttinger}},\
  }\bibfield  {title} {\enquote {\bibinfo {title} {Perfect entropy functions of
  the lattice boltzmann method},}\ }\href@noop {} {\bibfield  {journal}
  {\bibinfo  {journal} {EPL (Europhysics Letters)}\ }\textbf {\bibinfo {volume}
  {47}},\ \bibinfo {pages} {182} (\bibinfo {year} {1999})}\BibitemShut
  {NoStop}%
\bibitem [{\citenamefont {Lallemand}\ and\ \citenamefont
  {Luo}(2000)}]{lallemand2000theory}%
  \BibitemOpen
  \bibfield  {author} {\bibinfo {author} {\bibfnamefont {P.}~\bibnamefont
  {Lallemand}}\ and\ \bibinfo {author} {\bibfnamefont {L.-S.}\ \bibnamefont
  {Luo}},\ }\bibfield  {title} {\enquote {\bibinfo {title} {Theory of the
  lattice boltzmann method: Dispersion, dissipation, isotropy, galilean
  invariance, and stability},}\ }\href@noop {} {\bibfield  {journal} {\bibinfo
  {journal} {Physical review E}\ }\textbf {\bibinfo {volume} {61}},\ \bibinfo
  {pages} {6546} (\bibinfo {year} {2000})}\BibitemShut {NoStop}%
\bibitem [{\citenamefont {Teixeira}(1998)}]{teixeira1998incorporating}%
  \BibitemOpen
  \bibfield  {author} {\bibinfo {author} {\bibfnamefont {C.~M.}\ \bibnamefont
  {Teixeira}},\ }\bibfield  {title} {\enquote {\bibinfo {title} {Incorporating
  turbulence models into the lattice-boltzmann method},}\ }\href@noop {}
  {\bibfield  {journal} {\bibinfo  {journal} {International Journal of Modern
  Physics C}\ }\textbf {\bibinfo {volume} {9}},\ \bibinfo {pages} {1159--1175}
  (\bibinfo {year} {1998})}\BibitemShut {NoStop}%
\bibitem [{\citenamefont {Chen}\ and\ \citenamefont
  {Doolen}(1998)}]{chen1998lattice}%
  \BibitemOpen
  \bibfield  {author} {\bibinfo {author} {\bibfnamefont {S.}~\bibnamefont
  {Chen}}\ and\ \bibinfo {author} {\bibfnamefont {G.~D.}\ \bibnamefont
  {Doolen}},\ }\bibfield  {title} {\enquote {\bibinfo {title} {Lattice
  boltzmann method for fluid flows},}\ }\href@noop {} {\bibfield  {journal}
  {\bibinfo  {journal} {Annual review of fluid mechanics}\ }\textbf {\bibinfo
  {volume} {30}},\ \bibinfo {pages} {329--364} (\bibinfo {year}
  {1998})}\BibitemShut {NoStop}%
\bibitem [{\citenamefont {He}, \citenamefont {Chen},\ and\ \citenamefont
  {Zhang}(1999)}]{he1999lattice}%
  \BibitemOpen
  \bibfield  {author} {\bibinfo {author} {\bibfnamefont {X.}~\bibnamefont
  {He}}, \bibinfo {author} {\bibfnamefont {S.}~\bibnamefont {Chen}}, \ and\
  \bibinfo {author} {\bibfnamefont {R.}~\bibnamefont {Zhang}},\ }\bibfield
  {title} {\enquote {\bibinfo {title} {A lattice boltzmann scheme for
  incompressible multiphase flow and its application in simulation of
  rayleigh--taylor instability},}\ }\href@noop {} {\bibfield  {journal}
  {\bibinfo  {journal} {Journal of Computational Physics}\ }\textbf {\bibinfo
  {volume} {152}},\ \bibinfo {pages} {642--663} (\bibinfo {year}
  {1999})}\BibitemShut {NoStop}%
\bibitem [{\citenamefont {Liu}, \citenamefont {Valocchi},\ and\ \citenamefont
  {Kang}(2012)}]{liu2012three}%
  \BibitemOpen
  \bibfield  {author} {\bibinfo {author} {\bibfnamefont {H.}~\bibnamefont
  {Liu}}, \bibinfo {author} {\bibfnamefont {A.~J.}\ \bibnamefont {Valocchi}}, \
  and\ \bibinfo {author} {\bibfnamefont {Q.}~\bibnamefont {Kang}},\ }\bibfield
  {title} {\enquote {\bibinfo {title} {Three-dimensional lattice boltzmann
  model for immiscible two-phase flow simulations},}\ }\href@noop {} {\bibfield
   {journal} {\bibinfo  {journal} {Physical Review E}\ }\textbf {\bibinfo
  {volume} {85}},\ \bibinfo {pages} {046309} (\bibinfo {year}
  {2012})}\BibitemShut {NoStop}%
\bibitem [{\citenamefont {Reis}\ and\ \citenamefont
  {Phillips}(2007)}]{reis2007lattice}%
  \BibitemOpen
  \bibfield  {author} {\bibinfo {author} {\bibfnamefont {T.}~\bibnamefont
  {Reis}}\ and\ \bibinfo {author} {\bibfnamefont {T.}~\bibnamefont
  {Phillips}},\ }\bibfield  {title} {\enquote {\bibinfo {title} {Lattice
  boltzmann model for simulating immiscible two-phase flows},}\ }\href@noop {}
  {\bibfield  {journal} {\bibinfo  {journal} {Journal of Physics A:
  Mathematical and Theoretical}\ }\textbf {\bibinfo {volume} {40}},\ \bibinfo
  {pages} {4033} (\bibinfo {year} {2007})}\BibitemShut {NoStop}%
\bibitem [{\citenamefont {Chiappini}\ \emph {et~al.}(2019)\citenamefont
  {Chiappini}, \citenamefont {Sbragaglia}, \citenamefont {Xue},\ and\
  \citenamefont {Falcucci}}]{chiappini2019}%
  \BibitemOpen
  \bibfield  {author} {\bibinfo {author} {\bibfnamefont {D.}~\bibnamefont
  {Chiappini}}, \bibinfo {author} {\bibfnamefont {M.}~\bibnamefont
  {Sbragaglia}}, \bibinfo {author} {\bibfnamefont {X.}~\bibnamefont {Xue}}, \
  and\ \bibinfo {author} {\bibfnamefont {G.}~\bibnamefont {Falcucci}},\
  }\bibfield  {title} {\enquote {\bibinfo {title} {Hydrodynamic behavior of the
  pseudopotential lattice boltzmann method for interfacial flows},}\
  }\href@noop {} {\bibfield  {journal} {\bibinfo  {journal} {Phys. Rev. E}\
  }\textbf {\bibinfo {volume} {99}},\ \bibinfo {pages} {053305} (\bibinfo
  {year} {2019})}\BibitemShut {NoStop}%
\bibitem [{\citenamefont {Chiappini}\ \emph {et~al.}(2018)\citenamefont
  {Chiappini}, \citenamefont {Xue}, \citenamefont {Falcucci},\ and\
  \citenamefont {Sbragaglia}}]{chiappini2018ligament}%
  \BibitemOpen
  \bibfield  {author} {\bibinfo {author} {\bibfnamefont {D.}~\bibnamefont
  {Chiappini}}, \bibinfo {author} {\bibfnamefont {X.}~\bibnamefont {Xue}},
  \bibinfo {author} {\bibfnamefont {G.}~\bibnamefont {Falcucci}}, \ and\
  \bibinfo {author} {\bibfnamefont {M.}~\bibnamefont {Sbragaglia}},\ }\bibfield
   {title} {\enquote {\bibinfo {title} {Ligament break-up simulation through
  pseudo-potential lattice boltzmann method},}\ }in\ \href@noop {} {\emph
  {\bibinfo {booktitle} {AIP Conference Proceedings}}},\ Vol.\ \bibinfo
  {volume} {1978}\ (\bibinfo {organization} {AIP Publishing},\ \bibinfo {year}
  {2018})\ p.\ \bibinfo {pages} {420003}\BibitemShut {NoStop}%
\bibitem [{\citenamefont {Koda}\ and\ \citenamefont
  {Lien}(2015)}]{koda2015lattice}%
  \BibitemOpen
  \bibfield  {author} {\bibinfo {author} {\bibfnamefont {Y.}~\bibnamefont
  {Koda}}\ and\ \bibinfo {author} {\bibfnamefont {F.-S.}\ \bibnamefont
  {Lien}},\ }\bibfield  {title} {\enquote {\bibinfo {title} {The lattice
  boltzmann method implemented on the gpu to simulate the turbulent flow over a
  square cylinder confined in a channel},}\ }\href@noop {} {\bibfield
  {journal} {\bibinfo  {journal} {Flow, Turbulence and Combustion}\ }\textbf
  {\bibinfo {volume} {94}},\ \bibinfo {pages} {495--512} (\bibinfo {year}
  {2015})}\BibitemShut {NoStop}%
\bibitem [{\citenamefont {Fan}\ \emph {et~al.}(2021)\citenamefont {Fan},
  \citenamefont {Santasmasas}, \citenamefont {Guo}, \citenamefont {Yang},\ and\
  \citenamefont {Revell}}]{fan2021source}%
  \BibitemOpen
  \bibfield  {author} {\bibinfo {author} {\bibfnamefont {S.}~\bibnamefont
  {Fan}}, \bibinfo {author} {\bibfnamefont {M.~C.}\ \bibnamefont
  {Santasmasas}}, \bibinfo {author} {\bibfnamefont {X.-W.}\ \bibnamefont
  {Guo}}, \bibinfo {author} {\bibfnamefont {C.}~\bibnamefont {Yang}}, \ and\
  \bibinfo {author} {\bibfnamefont {A.}~\bibnamefont {Revell}},\ }\bibfield
  {title} {\enquote {\bibinfo {title} {Source term-based turbulent flow
  simulation on gpu with link-wise artificial compressibility method},}\
  }\href@noop {} {\bibfield  {journal} {\bibinfo  {journal} {International
  Journal of Computational Fluid Dynamics}\ }\textbf {\bibinfo {volume} {35}},\
  \bibinfo {pages} {549--561} (\bibinfo {year} {2021})}\BibitemShut {NoStop}%
\bibitem [{\citenamefont {Buffa}, \citenamefont {Jacob},\ and\ \citenamefont
  {Sagaut}(2021)}]{buffa2021lattice}%
  \BibitemOpen
  \bibfield  {author} {\bibinfo {author} {\bibfnamefont {E.}~\bibnamefont
  {Buffa}}, \bibinfo {author} {\bibfnamefont {J.}~\bibnamefont {Jacob}}, \ and\
  \bibinfo {author} {\bibfnamefont {P.}~\bibnamefont {Sagaut}},\ }\bibfield
  {title} {\enquote {\bibinfo {title} {Lattice-boltzmann-based large-eddy
  simulation of high-rise building aerodynamics with inlet turbulence
  reconstruction},}\ }\href@noop {} {\bibfield  {journal} {\bibinfo  {journal}
  {Journal of Wind Engineering and Industrial Aerodynamics}\ }\textbf {\bibinfo
  {volume} {212}},\ \bibinfo {pages} {104560} (\bibinfo {year}
  {2021})}\BibitemShut {NoStop}%
\bibitem [{\citenamefont {Guo}, \citenamefont {Zheng},\ and\ \citenamefont
  {Shi}(2002)}]{guo2002discrete}%
  \BibitemOpen
  \bibfield  {author} {\bibinfo {author} {\bibfnamefont {Z.}~\bibnamefont
  {Guo}}, \bibinfo {author} {\bibfnamefont {C.}~\bibnamefont {Zheng}}, \ and\
  \bibinfo {author} {\bibfnamefont {B.}~\bibnamefont {Shi}},\ }\bibfield
  {title} {\enquote {\bibinfo {title} {Discrete lattice effects on the forcing
  term in the lattice boltzmann method},}\ }\href@noop {} {\bibfield  {journal}
  {\bibinfo  {journal} {Physical review E}\ }\textbf {\bibinfo {volume} {65}},\
  \bibinfo {pages} {046308} (\bibinfo {year} {2002})}\BibitemShut {NoStop}%
\bibitem [{\citenamefont {Latt}(2007)}]{Latt2007}%
  \BibitemOpen
  \bibfield  {author} {\bibinfo {author} {\bibfnamefont {J.}~\bibnamefont
  {Latt}},\ }\emph {\bibinfo {title} {{Hydrodynamic limit of lattice Boltzmann
  equations}}},\ \href {\doibase 10.13097/archive-ouverte/unige:464} {Ph.D.
  thesis},\ \bibinfo  {school} {Univ. Geneve} (\bibinfo {year}
  {2007})\BibitemShut {NoStop}%
\bibitem [{\citenamefont {Davidson}\ \emph {et~al.}(2018)\citenamefont
  {Davidson} \emph {et~al.}}]{davidson2018fluid}%
  \BibitemOpen
  \bibfield  {author} {\bibinfo {author} {\bibfnamefont {L.}~\bibnamefont
  {Davidson}} \emph {et~al.},\ }\bibfield  {title} {\enquote {\bibinfo {title}
  {Fluid mechanics, turbulent flow and turbulence modeling},}\ }\href@noop {}
  {\bibfield  {journal} {\bibinfo  {journal} {Chalmers University of
  Technology, Goteborg, Sweden (Nov 2011)}\ } (\bibinfo {year}
  {2018})}\BibitemShut {NoStop}%
\bibitem [{\citenamefont {Kraichnan}(1970)}]{kraichnan1970diffusion}%
  \BibitemOpen
  \bibfield  {author} {\bibinfo {author} {\bibfnamefont {R.~H.}\ \bibnamefont
  {Kraichnan}},\ }\bibfield  {title} {\enquote {\bibinfo {title} {Diffusion by
  a random velocity field},}\ }\href@noop {} {\bibfield  {journal} {\bibinfo
  {journal} {The physics of fluids}\ }\textbf {\bibinfo {volume} {13}},\
  \bibinfo {pages} {22--31} (\bibinfo {year} {1970})}\BibitemShut {NoStop}%
\bibitem [{\citenamefont {Wilcox}(1988)}]{wilcox:88}%
  \BibitemOpen
  \bibfield  {author} {\bibinfo {author} {\bibfnamefont {D.~C.}\ \bibnamefont
  {Wilcox}},\ }\bibfield  {title} {\enquote {\bibinfo {title} {Reassessment of
  the scale-determining equation},}\ }\href@noop {} {\ \textbf {\bibinfo
  {volume} {26}},\ \bibinfo {pages} {1299--1310} (\bibinfo {year}
  {1988})}\BibitemShut {NoStop}%
\bibitem [{\citenamefont {Latt}\ \emph {et~al.}(2008)\citenamefont {Latt},
  \citenamefont {Chopard}, \citenamefont {Malaspinas}, \citenamefont
  {Deville},\ and\ \citenamefont {Michler}}]{latt2008straight}%
  \BibitemOpen
  \bibfield  {author} {\bibinfo {author} {\bibfnamefont {J.}~\bibnamefont
  {Latt}}, \bibinfo {author} {\bibfnamefont {B.}~\bibnamefont {Chopard}},
  \bibinfo {author} {\bibfnamefont {O.}~\bibnamefont {Malaspinas}}, \bibinfo
  {author} {\bibfnamefont {M.}~\bibnamefont {Deville}}, \ and\ \bibinfo
  {author} {\bibfnamefont {A.}~\bibnamefont {Michler}},\ }\bibfield  {title}
  {\enquote {\bibinfo {title} {Straight velocity boundaries in the lattice
  boltzmann method},}\ }\href@noop {} {\bibfield  {journal} {\bibinfo
  {journal} {Physical Review E}\ }\textbf {\bibinfo {volume} {77}},\ \bibinfo
  {pages} {056703} (\bibinfo {year} {2008})}\BibitemShut {NoStop}%
\bibitem [{\citenamefont {Zou}\ and\ \citenamefont
  {He}(1997)}]{zou1997pressure}%
  \BibitemOpen
  \bibfield  {author} {\bibinfo {author} {\bibfnamefont {Q.}~\bibnamefont
  {Zou}}\ and\ \bibinfo {author} {\bibfnamefont {X.}~\bibnamefont {He}},\
  }\bibfield  {title} {\enquote {\bibinfo {title} {On pressure and velocity
  boundary conditions for the lattice boltzmann bgk model},}\ }\href@noop {}
  {\bibfield  {journal} {\bibinfo  {journal} {Physics of fluids}\ }\textbf
  {\bibinfo {volume} {9}},\ \bibinfo {pages} {1591--1598} (\bibinfo {year}
  {1997})}\BibitemShut {NoStop}%
\bibitem [{\citenamefont {Carlsson}\ \emph {et~al.}(2022)\citenamefont
  {Carlsson}, \citenamefont {Davidson}, \citenamefont {Peng},\ and\
  \citenamefont {Arvidson}}]{carlsson2022investigation}%
  \BibitemOpen
  \bibfield  {author} {\bibinfo {author} {\bibfnamefont {M.}~\bibnamefont
  {Carlsson}}, \bibinfo {author} {\bibfnamefont {L.}~\bibnamefont {Davidson}},
  \bibinfo {author} {\bibfnamefont {S.-H.}\ \bibnamefont {Peng}}, \ and\
  \bibinfo {author} {\bibfnamefont {S.}~\bibnamefont {Arvidson}},\ }\bibfield
  {title} {\enquote {\bibinfo {title} {Investigation of turbulence injection
  methods in compressible flow solvers in large eddy simulation},}\ }in\
  \href@noop {} {\emph {\bibinfo {booktitle} {AIAA SCITECH 2022 Forum}}}\
  (\bibinfo {year} {2022})\ p.\ \bibinfo {pages} {0483}\BibitemShut {NoStop}%
\bibitem [{\citenamefont {Davidson}(2021)}]{pyCALC-RANS}%
  \BibitemOpen
  \bibfield  {author} {\bibinfo {author} {\bibfnamefont {L.}~\bibnamefont
  {Davidson}},\ }\href@noop {} {\enquote {\bibinfo {title} {{pyCALC-RANS:} a
  {2D} {P}ython code for {RANS}},}\ }\bibinfo {howpublished} {Division of Fluid
  Dynamics, Dept. of Mechanics and Maritime Sciences, Chalmers University of
  Technology{,} Gothenburg} (\bibinfo {year} {2021})\BibitemShut {NoStop}%
\bibitem [{\citenamefont {Guo}, \citenamefont {Kleiser},\ and\ \citenamefont
  {Adams}(1994)}]{guo1994comparison}%
  \BibitemOpen
  \bibfield  {author} {\bibinfo {author} {\bibfnamefont {Y.}~\bibnamefont
  {Guo}}, \bibinfo {author} {\bibfnamefont {L.}~\bibnamefont {Kleiser}}, \ and\
  \bibinfo {author} {\bibfnamefont {N.}~\bibnamefont {Adams}},\ }\bibfield
  {title} {\enquote {\bibinfo {title} {A comparison study of an improved
  temporal dns and spatial dns of compressible boundary layer transition},}\
  }in\ \href@noop {} {\emph {\bibinfo {booktitle} {Fluid Dynamics
  Conference}}}\ (\bibinfo {year} {1994})\ p.\ \bibinfo {pages}
  {2371}\BibitemShut {NoStop}%
\bibitem [{\citenamefont {Adams}(1998)}]{adams1998direct}%
  \BibitemOpen
  \bibfield  {author} {\bibinfo {author} {\bibfnamefont {N.~A.}\ \bibnamefont
  {Adams}},\ }\bibfield  {title} {\enquote {\bibinfo {title} {Direct numerical
  simulation of turbulent compression ramp flow},}\ }\href@noop {} {\bibfield
  {journal} {\bibinfo  {journal} {Theoretical and Computational Fluid
  Dynamics}\ }\textbf {\bibinfo {volume} {12}},\ \bibinfo {pages} {109--129}
  (\bibinfo {year} {1998})}\BibitemShut {NoStop}%
\bibitem [{\citenamefont {Moser}, \citenamefont {Kim},\ and\ \citenamefont
  {Mansour}(1999)}]{moser1999direct}%
  \BibitemOpen
  \bibfield  {author} {\bibinfo {author} {\bibfnamefont {R.~D.}\ \bibnamefont
  {Moser}}, \bibinfo {author} {\bibfnamefont {J.}~\bibnamefont {Kim}}, \ and\
  \bibinfo {author} {\bibfnamefont {N.~N.}\ \bibnamefont {Mansour}},\
  }\bibfield  {title} {\enquote {\bibinfo {title} {Direct numerical simulation
  of turbulent channel flow up to re $\tau$= 590},}\ }\href@noop {} {\bibfield
  {journal} {\bibinfo  {journal} {Physics of fluids}\ }\textbf {\bibinfo
  {volume} {11}},\ \bibinfo {pages} {943--945} (\bibinfo {year}
  {1999})}\BibitemShut {NoStop}%
\bibitem [{\citenamefont {Favier}, \citenamefont {Godeferd},\ and\
  \citenamefont {Cambon}(2010)}]{favier2010space}%
  \BibitemOpen
  \bibfield  {author} {\bibinfo {author} {\bibfnamefont {B.}~\bibnamefont
  {Favier}}, \bibinfo {author} {\bibfnamefont {F.}~\bibnamefont {Godeferd}}, \
  and\ \bibinfo {author} {\bibfnamefont {C.}~\bibnamefont {Cambon}},\
  }\bibfield  {title} {\enquote {\bibinfo {title} {On space and time
  correlations of isotropic and rotating turbulence},}\ }\href@noop {}
  {\bibfield  {journal} {\bibinfo  {journal} {Physics of Fluids}\ }\textbf
  {\bibinfo {volume} {22}},\ \bibinfo {pages} {015101} (\bibinfo {year}
  {2010})}\BibitemShut {NoStop}%
\end{thebibliography}%

\end{document}